\documentclass[twocolumn]{aastex61}
\usepackage{longtable}
\usepackage{float}
\usepackage{threeparttablex, tablefootnote}

\submitjournal{ApJ}

\shorttitle{Magnetic Fields of M Dwarfs from the Pleiades Open Cluster}
\shortauthors{Wanderley et al.}

\begin{document}

\title{Magnetic Fields in M Dwarf Members of the Pleiades Open Cluster Using APOGEE spectra}
\correspondingauthor{Fábio Wanderley}
\email{fabiowanderley@on.br}

\author[0000-0003-0697-2209]{Fábio Wanderley}
\affiliation{Observatório Nacional/MCTIC, R. Gen. José Cristino, 77, 20921-400, Rio de Janeiro, Brazil}

\author[0000-0001-6476-0576]{Katia Cunha}
\affiliation{Observatório Nacional/MCTIC, R. Gen. José Cristino, 77,  20921-400, Rio de Janeiro, Brazil}
\affiliation{Steward Observatory, University of Arizona, 933 North Cherry Avenue, Tucson, AZ 85721-0065, USA}

\author[0000-0003-3061-4591]{Oleg Kochukhov}
\affiliation{Department of Physics and Astronomy, Uppsala University, Box 516, S-75120 Uppsala, Sweden}

\author[0000-0002-0134-2024]{Verne V. Smith}
\affiliation{NSF’s NOIRLab, 950 N. Cherry Ave. Tucson, AZ 85719 USA}

\author[0000-0002-7883-5425]{Diogo Souto}
\affiliation{Departamento de F\'isica, Universidade Federal de Sergipe, Av. Marcelo Deda Chagas, S/N Cep 49.107-230, S\~ao Crist\'ov\~ao, SE, Brazil}

\author[0000-0002-8849-9816]{Lyra Cao}
\affiliation{Department of Astronomy, The Ohio State University, Columbus, OH 43210, USA}

\author[0000-0001-6914-7797]{Kevin Covey}
\affiliation{Department of Physics $\&$ Astronomy, Western Washington University, Bellingham, WA, 98225, USA}

\author[0000-0003-2025-3147]{Steven R. Majewski}
\affiliation{Department of Astronomy, University of Virginia, Charlottesville, VA 22904-4325, USA}

\author[0000-0003-2745-8241]{Cintia Martinez}
\affiliation{Instituto de Astronomía y Física del Espacio (CONICET-UBA), C.C. 67 Sucursal 28, C1428EHA, Buenos Aires, Argentina}

\author[0000-0002-0638-8822]{Philip S. Muirhead}
\affiliation{Institute for Astrophysical Research and Department of Astronomy, Boston University, 725 Commonwealth Ave., Boston, MA 02215, USA}
\affiliation{Center for Interdisciplinary Exploration and Research in Astrophysics (CIERA) and Department of Physics and Astronomy, Northwestern University, 1800 Sherman Ave., Evanston, IL 60201, USA}

\author[0000-0002-7549-7766]{Marc Pinsonneault}
\affiliation{Department of Astronomy, The Ohio State University, Columbus, OH 43210, USA}

\author[0000-0002-0084-572X]{C. Allende Prieto}
\affiliation{Instituto de Astrofísica de Canarias, E-38205 La Laguna, Tenerife, Spain}
\affiliation{Departamento de Astrofísica, Universidad de La Laguna, E-38206 La Laguna, Tenerife, Spain}

\author[0000-0002-3481-9052]{Keivan G. Stassun}
\affiliation{Department of Physics and Astronomy, Vanderbilt University, 6301 Stevenson Center Ln., Nashville, TN 37235, USA}

\begin{abstract}
Average magnetic field measurements are presented for 62 M-dwarf members of the Pleiades open cluster, derived from Zeeman-enhanced Fe I lines in the H-band. An MCMC methodology was employed to model magnetic filling factors using SDSS-IV APOGEE high-resolution spectra, along with the radiative transfer code SYNMAST, MARCS stellar atmosphere models, and the APOGEE DR17 spectral line list. 
There is a positive correlation between mean magnetic fields and stellar rotation, with slow-rotator stars (Rossby number, Ro$>$0.13) exhibiting a steeper slope than rapid-rotators (Ro$<$0.13). However, the latter sample still shows a positive trend between Ro and magnetic fields, which is given by 
$<$B$>$ = 1604 $\times$ Ro$^{-0.20}$.
The derived stellar radii, when compared with physical isochrones, show that on average, our sample shows radius inflation, with median enhanced radii ranging from 
+3.0$\%$ to +7.0$\%$, depending on the model.  There is a positive correlation between magnetic field strength and radius inflation, as well as with stellar spot coverage, correlations that together indicate that stellar spot-filling factors generated by strong magnetic fields might be the mechanism that drives radius inflation in these stars. We also compare our derived magnetic fields with chromospheric emission lines (H$\alpha$, H$\beta$ and Ca II K), as well as with X-ray and H$\alpha$ to bolometric luminosity ratios, and find that stars with higher chromospheric and coronal activity tend to be more magnetic.
\end{abstract}


\section{Introduction}

Quantitative characterization of magnetic fields provides
a deeper understanding of stellar physics. 
As a star evolves, its stellar wind interacts with the magnetosphere, generating a torque that converts kinetic energy into magnetic energy, reducing the stellar angular momentum and resulting in a slowing of its rotational velocity \citep{kawaler1988}.  
This process of magnetic braking over time enables gyrochronology to estimate the age of a star based on its stellar rotation \citep{skumanich1972,barnes2003}, with younger stars tending to have higher magnetic fields and activity than older stars of similar T$_{\rm eff}$. Another effect caused by magnetic fields is the heating of the stellar chromosphere and coronae, causing the emission of, respectively, strong UV and X-ray non-thermal radiation \citep{hawley2014,astudillo2017,newton2017}.

Magnetic fields are especially important in M dwarf stars, as these stars have longer spin-down timescales, maintaining magnetic fields for longer periods than hotter stars \citep{newton2016}. This means that any surrounding exoplanets will experience high-energy fluxes for longer periods, which can impact the habitability of these systems. In addition, since M dwarfs are cool, their habitable zones are located closer when compared to hotter stars, which increases the incident flux, as well as the probability of orbit locking, which, in turn, also impacts habitability. Nonetheless, M dwarf stars represent around $70 \%$ of the stars of our galaxy \citep{salpeter1955,reid1997}, and some of them can live trillions of years on the main sequence, which gives life many opportunities and time to form and evolve around these stars. A deep understanding of their magnetic fields and implications for their environment is fundamental for the understanding of these stars and the habitability of their exo-planetary systems.

One method to study stellar magnetic fields is through the Zeeman effect. Magnetically sensitive lines (e.g., having high effective Landé g-factors) when subject to magnetic fields are split into components, which we can observe as an additional line broadening. This broadening scales with the square of the line central wavelength, therefore spectral lines located at longer wavelengths are more sensitive to Zeeman splitting than lines with the same effective Landé-g factors located in the bluer part of the spectrum (see \citealt{kochukhov2021} and references therein)

Due to other broadening mechanisms, such as Doppler broadening, stellar rotation, and instrumental broadening, we may not be able to resolve the Zeeman splitting in the spectrum, and what we measure is the Zeeman intensification of an affected line \citep{stift2003,basri1992,basri1994}.


There are two main ways to characterize stellar magnetic fields, one considering large-scale and the other small-scale magnetic fields. Large-scale analyses provide a topological view of the stellar magnetic field, separating it into components based on different orientations. The technique used for this characterization, Zeeman-Doppler Imaging (ZDI, \citealt{kochukhov2016}) is based on the analysis of the circular polarization of the line, described by the Stokes $V$ parameter, and therefore spectro-polarimetric data are needed for this technique. The small-scale magnetic approach models the total intensity of the field and is based on the Stokes $I$ parameter, and no polarimetric stellar data are needed. 
The review by \citet{kochukhov2021} discusses the current state of M dwarf magnetic field studies and presents a compilation of large and small-scale magnetic field measurements from the literature. Below we mention the results of a few of these studies.

The first work to model the magnetic field for an M dwarf star in the literature was \citet{saar1985_olegreview11}, who used Ti I lines from a high-resolution (R$\sim$45,000) K-band spectrum of the flare star AD Leo and found a mean magnetic field of 3.8 kG. Many works followed, modeling the Zeeman effect in M dwarfs spectral lines in the optical and infrared and finding magnetic fields ranging from zero up to 8 kG \citep{johns-krull1996_olegreview15, shulyak2011_olegreview1,shulyak2014_olegreview10,shulyak2017_olegreview2,shulyak2019_olegreview3, reiners2022, cristofari2023a, cristofari2023b, eunkyu2023}. \citet{reiners2022} (hereafter, R22) determined magnetic fields for a large sample of M dwarfs using CARMENES spectra, and analyzed how magnetic fields are related to parameters such as magnetic flux, activity, and Rossby numbers, and how these distributions change when comparing the saturated and non-saturated regimes. Some works studied magnetic fields deriving large-scale magnetic fields from Stokes V, as well as small-scale fields, and found that the latter is considerably greater than the one obtained from circular polarization, which indicates that most of the star's magnetic field is probably stored in small structures at its surface \citep{phan-bao_olegreview13, kochukhov2017_olegreview4, kochukhov2019_olegreview7, kochukhov2020}.

All of the studies mentioned above explored magnetic fields in M dwarfs from the Galactic field star population and these can in principle have different ages and metallicities. Stars from an open cluster, on the contrary, originate from the same molecular cloud and are expected to have approximately the same age and chemistry, making them great benchmarks with which to study stellar evolution and atomic diffusion, but these are also excellent benchmarks for studying stellar magnetic fields. Because cluster stars form at the same time and share the same chemical composition, metallicity and age dependencies are removed by their inter-comparison, and this allows for an investigation of magnetic fields primarily as a function of other stellar properties, such as effective temperatures or rotational periods \citep{souto2021}. The recent work of \citet{wanderley2023} used APOGEE near-infrared spectra \citep{majewski2017_apogee,apogeedr17_2022} to derive atmospheric parameters and metallicities, and study radius inflation in a sample of M dwarf stars members of the young Hyades open cluster. \citep{wanderley2023} found that these stars are on average inflated, and this may be caused by stellar magnetic fields.

In this work, we use spectral lines affected by Zeeman broadening present in the SDSS APOGEE spectra to derive average magnetic fields for a sample of 62 M dwarfs from the young (age = 112 $\pm$ 5 Myr; \citealt{dahm2015_age4}), near-solar metallicity \citep{soderblom2009} Pleiades open cluster. 
This is the first study to derive magnetic fields for a sample of M dwarf members of an open cluster, and also the first to derive magnetic fields for M dwarfs based on APOGEE H-band spectra.

This paper is organized as follows: Section 2 presents the APOGEE data and sample selection. In Section 3 we present the methodology employed to derive average stellar magnetic fields for the Pleiades M dwarf star sample. In Section 4, we discuss the results which include the relation between magnetic fields and stellar rotation, comparisons with the literature, radius inflation, and analysis of activity indicators. Finally, Section 5 summarizes the conclusions.

\section{APOGEE Data and Sample Selection}

We determined average magnetic fields by analyzing near-infrared ($\lambda$1.51$\mu$m to $\lambda$1.69$\mu$m), high-resolution (average resolution of $R\sim22,500$) spectra of Pleiades M dwarf stars observed by the SDSS IV APOGEE survey \citep{majewski2017_apogee,blanton2017}. As part of SDSS IV, the APOGEE spectra analyzed here were obtained at the 2.5-m telescope located at APO in the northern hemisphere \citep{bowen1973,gunn2006_sdss,wilson2019}. 

An initial sample of Pleiades members was obtained from \citet{HEYL_2022} and confirmed using the membership analysis in \citet{CANTATGAUDIN_2020}. We adopted a threshold of $80\%$ of minimum membership probability from \citet{CANTATGAUDIN_2020} for a star to be considered a member of the Pleiades.
We cross matched this sample with APOGEE DR17 \citep{apogeedr17_2022}, and selected for M dwarfs with $4.7 < M_{K_{s}} < 6.2$ \citep{mann2015,mann2016}, using 2MASS $K_{S}$ magnitudes \citep{skrustkie2006} and distances from \citet{bailerjones2021}. All magnitudes were corrected for extinction using the mean extinction to the Pleiades of A$_{\rm V}$ = 0.12 \citep{stauffer2007_vmag}, and the relations from \citet{wang2019}. 
To remove binary stars from the sample, we considered only stars with a scatter in APOGEE radial velocity smaller than 1 km s$^{-1}$. We also removed stars that presented large Gaia DR3 \citep{gaia2022} RUWE numbers (RUWE $>$ 1.4), as this can indicate the presence of an unresolved companion \citep{belokurov2020}. In addition, stars without a v$\sin{i}$ measurement in DR17 were also removed from the sample.

We also analyzed the distribution of distances, proper motions (Gaia DR3, \citealt{gaia2022}), and radial velocities (from APOGEE DR17) of the selected targets to check for outliers, but we found none. Finally, we also removed from the sample stars that had noisy and problematic APOGEE spectra. Our final sample of Pleiades members analyzed in this study is composed of 62 M dwarfs.

\begin{figure*}
\begin{center}
  \includegraphics[angle=0,width=0.8\linewidth,clip]{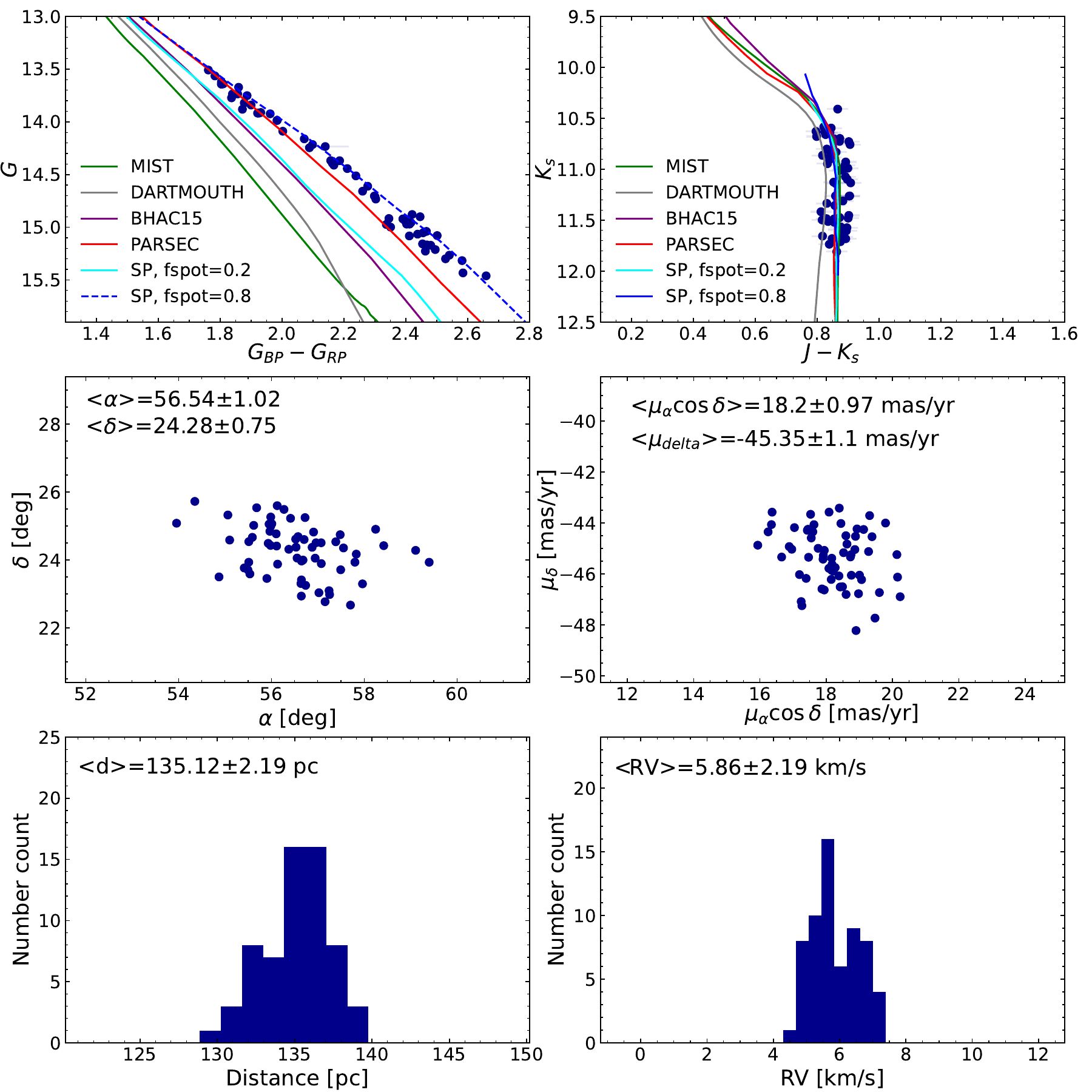}
\caption{From left to right, the top panels present respectively Gaia and 2MASS CMDs, the blue points are the selected M dwarf members of the Pleiades open cluster. Several 100 Myr solar metallicity isochrones are shown: MIST, DARTMOUTH, PARSEC, BHAC15, and SPOTS. Two SPOTS isochrones are shown, one for a photosphere spot coverage of 20$\%$ and another for 80$\%$. The middle panels present respectively the distribution in space (right ascension and declination) and proper motions (from Gaia DR3) of the target stars. The bottom panels present respectively the distance \citep{bailerjones2021} and radial velocity histograms. The middle and bottom panels also present the mean and standard deviations for the parameters.}
\end{center}
\label{membership}
\end{figure*}

Figure \ref{membership} presents the Gaia and 2MASS CMDs (color-magnitude diagram) of the selected targets (top panels), their distribution in space, and proper motions (middle panels), as well as histograms for their star distances ($d$), and radial velocities ($RV$) (bottom panels). The average distance, radial velocity, and proper motions along $\alpha$ (right ascension) and $\delta$ (declination) for our Pleiades M dwarf sample are, respectively: $<$d$>$=135.12$\pm$2.19 pc, $<$RV$>$=5.86$\pm$2.19 km s$^{-1}$, $<$ $\mu_{\alpha}\cos{\delta}$ $>$=18.20$\pm$0.97 mas yr$^{-1}$ and $<$ $\mu_{\delta}$ $>$=$-45.35\pm$1.10 mas yr$^{-1}$. These results are in good agreement with measurements from \citet{lodieu2019} of respectively: 135.15$\pm$0.43 pc, 5.67 $\pm$ 2.93 $km \, s^{-1}$, 19.5 mas yr$^{-1}$ and $-45.5$ mas yr$^{-1}$. 

In the top panels of Figure \ref{membership} we show several isochrones from the literature: a MIST isochrone \citep{choi2016_MIST}, a DARTMOUTH isochrone \citep{Dotter2008_dartmouth}, a PARSEC isochrone \citep{Bressan2012_parsec,nguyen2022}, a BHAC15 isochrone \citep{baraffe2015}, and two SPOTS isochrones from \cite{Somers2020_spots}, one with spots covering 20$\%$ of the stellar photosphere and another with a spot coverage of 80$\%$. All isochrones shown are for solar metallicity and an age of 100 Myr, which is roughly the estimated age for the Pleiades open cluster.  

The Pleiades study by \citet{covey2016} found significant scatter in the K vs J$-$K$_{s}$ diagram of Pleiades stars (see Figure 3 in their paper). That study also compared the observed colors of Pleiades stars with physical models and found that the V$-$K colors in rapidly rotating stars present a positive offset for the same V magnitude if compared to slow rotators, which was interpreted as being due to binarity or a dependency of photospheric/spot properties on rotation rate. 
The photometric data in the Gaia G vs G$_{\rm BP}-$G$_{\rm RP}$ CMD, shown in the top left panel of Figure \ref{membership}, 
shows a clear offset, with 
most isochrones that do not consider stellar spots presenting bluer colors for the same magnitudes, while the SPOTS isochrone associated with an 80$\%$ photospheric spot coverage presents an excellent match to the photometric data of the selected stars. 
We note that this pattern is not seen in the 2MASS CMD, where all isochrones present very small variations, even for different spot fractions, which might be related to the lower photometric spot contrast in the infrared when compared to the visible spectrum.


\section{Methodology and Results}

To derive average magnetic fields from Zeeman intensified lines we used the APOGEE line list \citep{smith2021} and model atmospheres from the MARCS grid \citep{gustafsson2008_marcs}.
The average magnetic field modeling in this study was done using the SYNMAST spectral synthesis code \citep{kochukhov2010_synmast}, which computes the effects of magnetic fields on stellar spectra. This code uses polarised radiative transfer calculations to derive IQUV local Stokes parameters for a given magnetic field vector (in radial, meridional, and azimuthal orientations). In this study, we assumed a radial magnetic field and used SYNMAST to calculate intensity at seven limb angles. Then another code was used to perform disk integration, converting these intensity fluxes into density fluxes, which can be compared to APOGEE spectra.

To search for the best iron lines in the APOGEE region that can be used as magnetic field indicators for M dwarf stars, we compared two SYNMAST syntheses, one computed for 0 kG (no magnetic field) and another for 3 kG. We selected four Fe I lines as best indicators: $\lambda$15207.526 \r{A}, $\lambda$15294.56 \r{A}, $\lambda$15621.654 \r{A}, and $\lambda$15631.948 \r{A}. 
Table \ref{linelist} presents the selected spectral lines, their central wavelengths in vacuum, the excitation potentials, $\log{\rm gf}$ values from the APOGEE line list \citep{smith2021}, effective Landé-g factors collected from VALD database \citep{piskunov1995,kupka1999}, as well as term designations associated with the upper and lower energy levels. The four selected Fe I lines are considerably sensitive to magnetic fields presenting effective Landé-g factors that range between $\sim1.5$ and $\sim1.7$. All the selected Fe I lines, except $\lambda$15621.654 \r{A} are from the same multiple. These lines were used to compute the magnetic fields for all stars in our sample and gave overall consistent results. 

After the selection of diagnostic lines for measuring magnetic fields, the next step in our analysis was to generate a grid of synthetic spectra which was used in the analysis of each star. We adopted the DR17 ASPCAP (APOGEE Stellar Parameter and Chemical Abundances Pipeline, \citealt{garciaperez2016_aspcap,apogeedr17_2022}) T$_{\rm eff}$ values for each star, along with an approximate $\log{g}$ ($\sim$ 4.7 -- 4.8), depending on the stellar T$_{\rm eff}$. The ASPCAP results used in this work were computed with the Turbospectrum \citep{plez2012_turbospectrum} code instead of Synspec \citep{hubeny2011_synspec}, however, we note that there are no significant differences between both sets of results for M dwarfs. 
The spectra of M dwarfs are relatively insensitive to the microturbulent velocity parameter, as noted by \citet{souto2017,souto2020}, who found that a value of 1 km s$^{-1}$ provides good fits to the observations, and we adopt this value in our analysis. 
All measurable Fe I lines in the APOGEE spectra of M dwarfs have high effective Landé-g factors, which makes them not suitable to be used as rotational broadening indicators. Therefore, we used a sample of OH lines, which are insensitive to magnetic fields.
This approach is similar to the one adopted by \citet{johns-krull2004,johns-krull2007,yang2008,yang2011,lavail2019}, who derived mean magnetic fields for T-Tauri stars from K-band near-infrared spectra, using magnetically insensitive CO lines to measure non-magnetic broadening.
To derive projected rotational velocities (v$\sin{i}$) for the stars we used the radiative transfer code Turbospectrum \citep{plez2012_turbospectrum}; we adopted v${\sin{i}}$ threshold of 3 km s$^{-1}$ given the spectral resolution of the APOGEE spectra. We computed a grid of synthetic spectra, for the adopted stellar parameters for each star and Fe I line, with metallicities ranging from -0.75$\leq$[Fe/H]$\leq$+0.5 in steps of 0.25, and magnetic field values from 0 to 12 kG in steps of 2 kG, considering only the radial component.
We then convolved the synthetic spectra with a rotational profile for the adopted v$\sin{i}$ as well as a Gaussian profile corresponding to the spectrum LSF (see \citealt{wilson2019,nidever2015}). 
Each synthesis was fitted to the DR17 normalized APOGEE spectrum \citep{garciaperez2016_aspcap} and was subject to small wavelength shifts when needed.

We employed Monte Carlo and Markov Chain (MCMC) to model the observed spectra and derive magnetic fields for the Pleiades M dwarfs. MCMC is a powerful tool, not only because it provides 
best fits to observations but also because it gives realistic and well-defined uncertainties based on the posterior distribution. Our methodology considers that the surface of the star can be divided into different components, each associated with a different magnetic field value. The filling factor describes the fraction of the stellar surface associated with a specific $<$B$>$. Many works in the literature used MCMC to derive average magnetic fields from filling factor determinations \citep{lavail2019,kochukhov2020,hahlin2021,hahlin2022,reiners2022,cristofari2023a,cristofari2023b,hahlin2023,pouilly2023}.

We developed a methodology to derive magnetic fields, that employs the python-code emcee \citep{emcee2013} and finds the combination of metallicity, and 6 filling factors (2--12 kG in a 2 kG step) that best fits the four selected Fe I lines at the same time. We note that the non-magnetic filling factor f$_{0}$ is given by 1$-\sum f_{n}$. For each entry in the posterior distribution, we calculate an average magnetic field in Gauss units given by: 

\begin{equation}
    <B>= \sum_{n} f_{n} \times 1000n \, \, , \, \,
    n=[2,4,6,8,10,12]
    \label{eqb}    
\end{equation}

The adopted average magnetic field, along with the lower and upper uncertainties, are given respectively by the median, the $16^{th}$, and $84^{th}$ percentiles of the posterior distribution. In Table \ref{filfactor} we provide the filling factors and mean magnetic fields for two stars, the ones having the highest and lowest $<$B$>$ in our sample.

\begin{figure*}
\begin{center}
  \includegraphics[angle=0,width=0.33\linewidth,clip]{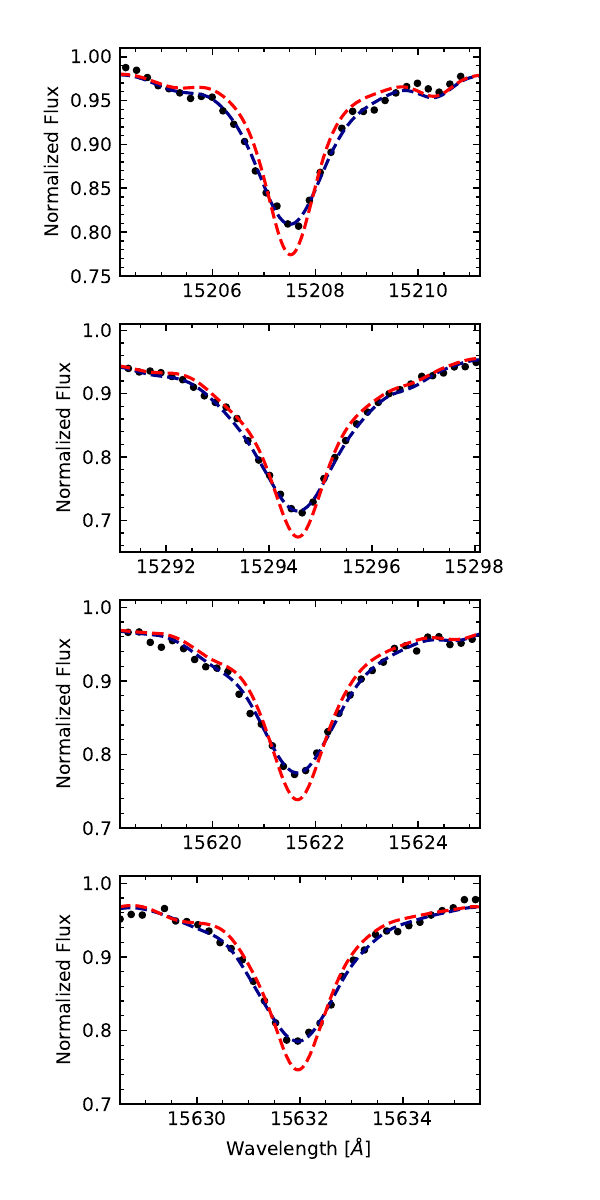}
  \includegraphics[angle=0,width=0.66\linewidth,clip]{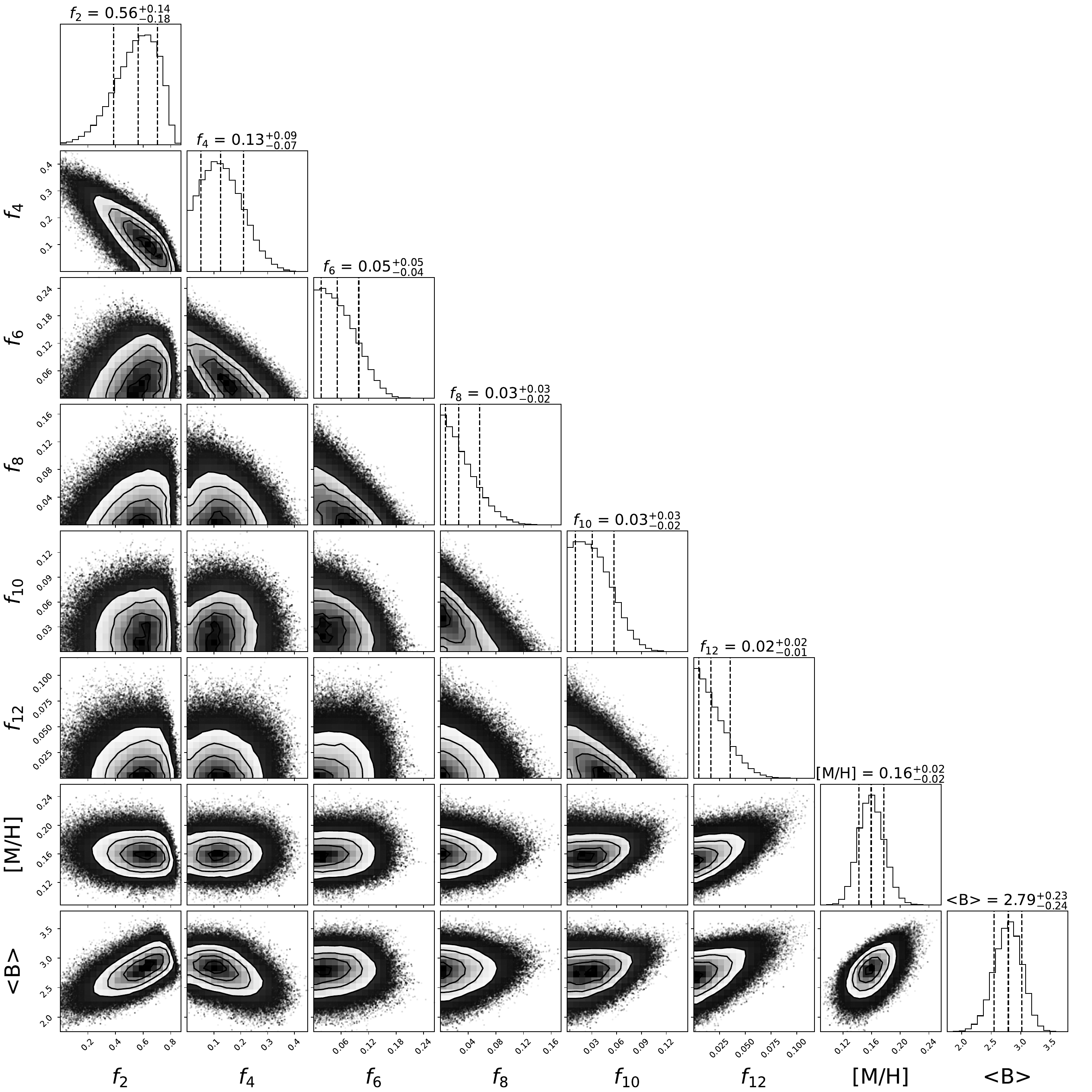}  
\caption{Mean magnetic field results for star 2M03511207+2355575. The left panels show the four Fe I lines used in the modeling: black dots are the observed APOGEE spectrum, red dashed lines are synthetic profiles computed without magnetic field, and dark blue lines are our best-fits obtained from the MCMC modeling. The right panel is a corner plot that presents the median and uncertainties (from $16^{th}$ and $84^{th}$ percentiles) of the derived parameters, which includes filling factors from 2 to 12 kG in steps of 2 kG. We also show the obtained average magnetic field (in kG) and its uncertainties.} 
\end{center}
\label{fitexample}
\end{figure*}

In Figure \ref{fitexample}, we illustrate the methodology by presenting the best fits, as well as the corner plot for the star 2M03511207+2355575. The left panels show fits for the four lines, where black dots are the observed spectrum, and the blue and red lines represent respectively the synthesis with the derived magnetic field, and the result with the same metallicity but no magnetic field. The right panel shows the corner plot describing the results. It presents the median and uncertainties (from $16^{th}$ and $84^{th}$ percentiles) of the modeled filling factors and the metallicity. We also show the posterior distribution for the derived average magnetic field. 

The derived mean magnetic fields for the studied Pleiades M dwarf stars range from $\sim$1.0 to $\sim$4.2 kG, with a median$\pm$MAD of 3.0$\pm$0.3 kG.
Table \ref{compiledata} presents the derived mean magnetic fields for the sample stars, along with the adopted T$_{\rm eff}$ and v$\sin{i}$, and the SNR of the APOGEE spectra analyzed. This table also contains some quantities that will be discussed in Section 4: rotational periods, and Rossby numbers the ratio between the derived magnetic fields and the magnetic field limit based on kinetic-to-magnetic energy conversion, magnetic fluxes, and activity indicators: X-ray to bolometric ratios and H$\alpha$, H$\beta$ and Ca II K equivalent widths.

\section{Discussion}

\subsection{Magnetic Fields, Rossby Numbers and Stellar Rotation}

We collected rotational periods for 53 stars in our sample from the works by \citet{hartman2010}, \citet{rebull2016} and \citet{covey2016}, who analyzed data respectively from the HATNet \citep{bakos2004}, K2 \citep{howell2014}, and PTF \citep{law2009,rau2009} surveys. The rotational periods of the sample stars include both slow and rapid rotators, encompassing rotational periods $\sim$0.7 to 17.3 days.

The Rossby number (Ro) is an important indicator of stellar activity, it is given by the ratio between the rotational period and the convective turnover time.
The convective turnover time, $\tau$, is defined as the time that it takes a convective cell to traverse the convective envelope of a star. Due to both their deeper convective envelopes and slower convective velocities, cooler stars are expected to have longer convective turnover times, and also to be able to sustain dynamos longer than hotter stars.
For $\tau$ measured in units of days, we used the relation $\tau=12.3\times (L/L_{\odot})^{-0.5}$, derived by R22. 
Bolometric luminosities were determined using distances from \citet{bailerjones2021}, 2MASS $K_{s}$ magnitudes, and bolometric corrections from photometric calibrations from \citet{mann2015,mann2016}. This calibration is based on 2MASS $J$ and $V$ band magnitudes collected from \citet{stauffer2007_vmag,lasker2008_gsc,zacharias2012_vmag,muirhead2018_vmag}. We also employed the same method to account for extinction described in Section 2. 

\begin{figure}[h!]
\begin{center}  \includegraphics[angle=0,width=1\linewidth,clip]{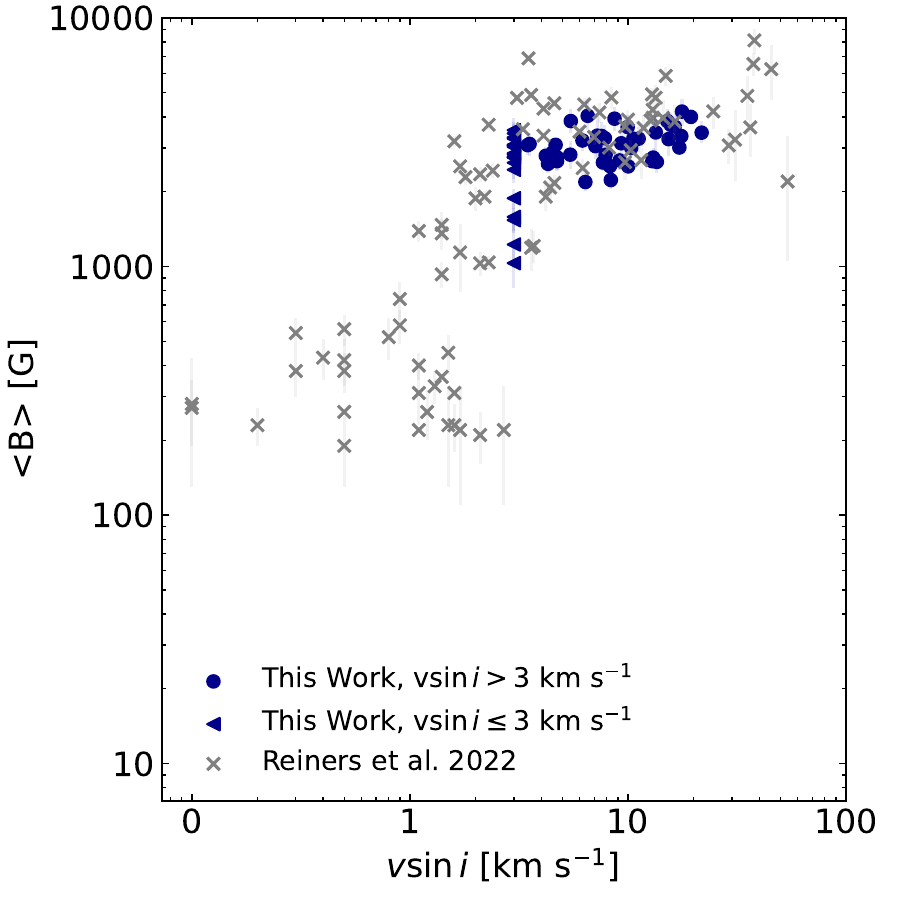}
\caption{Distribution of the derived average magnetic fields versus adopted projected rotational velocities, v$\sin{i}$, for the studied M dwarfs in the Pleiades are represented by blue circles and triangles for respectively stars with v$\sin{i}>3$ km s$^{-1}$ and with v$\sin{i} \leq 3$ km s$^{-1}$. Results from \citet{reiners2022} are also shown (black x's).}
\end{center}
\label{vsini_b}
\end{figure}

The mean magnetic fields derived for the sample stars versus their projected rotational velocities, v$\sin{i}$ (both of which are quantities derived from the APOGEE spectra), are shown in Figure \ref{vsini_b} as blue symbols. The blue circles represent stars for which we can estimate the v$\sin{i}$, while the blue triangles are stars having v$\sin{i}$ values up to 3 km s$^{-1}$, the adopted threshold in v$\sin{i}$ that can be estimated in this study. We also show in this figure literature results from R22 as black x's. The mean magnetic fields for the Pleiades stars generally overlap with the M dwarfs in R22 within the overlapping v$\sin{i}$ range, showing just a small systematic difference in $<$B$>$ when compared to R22. 

\begin{figure}[h!]
\begin{center}
  \includegraphics[angle=0,width=1\linewidth,clip]{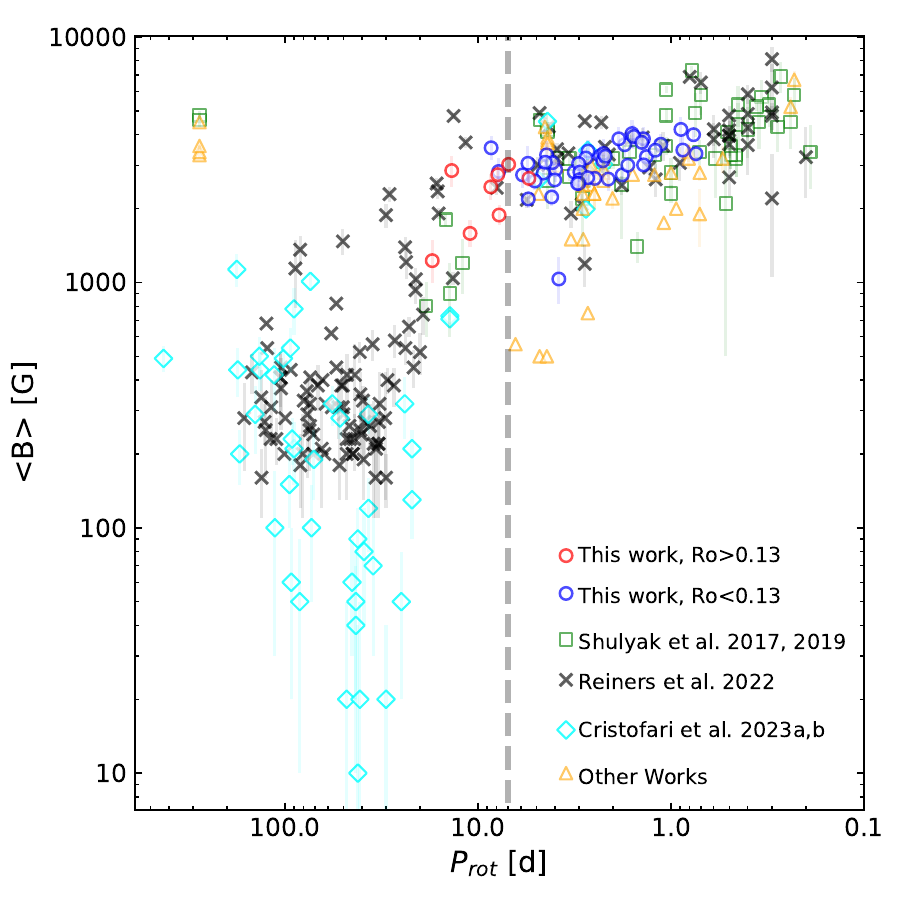}
\caption{Mean magnetic fields versus rotational periods for Pleiades M dwarfs rapid (open blue circles, Ro$<$0.13) and slow rotators (open red circles, Ro$>$0.13), along with data compiled from the literature. Green squares, black x's, and cyan diamonds are data from respectively \citet{shulyak2017_olegreview2,shulyak2019_olegreview3}, \citet{reiners2022}, and \citet{cristofari2023a,cristofari2023b}. Orange triangles show results from other works, see text for references. 
}
\end{center}
\label{litcomp}
\end{figure}

Figure \ref{litcomp} shows our results of magnetic fields versus rotational periods for the Pleiades M dwarfs (red and blue open circles), along with results from the literature for comparison. We compiled a total of 281 average magnetic field measurements from the literature. The results from \citet{shulyak2017_olegreview2} and \citet{shulyak2019_olegreview3} are represented by green squares in Figure \ref{litcomp}. Results from \citet{cristofari2023a} and \citet{cristofari2023b}, (results from radial-tangential macroturbulent velocity profiles and fitted $\log{g}$) are represented by cyan diamonds. Results from R22 are for stars with $T_{\rm eff}<4000$ K and are represented by black x's. Finally, other results from the literature (orange triangles), are from the following works: 
\citet{shulyak2011_olegreview1, kochukhov2017_olegreview4, kochukhov2001_olegreview5, kochukhov2009_olegreview6, kochukhov2019_olegreview7, johns-krull2000_olegreview8, afram2009_olegreview9, shulyak2014_olegreview10, saar1985_olegreview11, saar1994_olegreview12, phan-bao_olegreview13, saar1996_olegreview14, johns-krull1996_olegreview15}. 

We can divide the $<$B$>$--P$_{\rm rot}$ plane in Figure \ref{litcomp} into two regions that correspond to the saturated and unsaturated regimes. 
Red and blue open circles in this figure represent respectively stars in our sample that are slow rotators (Ro$>$0.13, non-saturated regime) and rapid rotators (Ro$<$0.13, saturated regime). 
We added a vertical dashed line at $P_{\rm rot}$=7 days to Figure \ref{litcomp}, which is an estimated threshold analogous to the separation based on Rossby numbers.
Stars in the non-saturated regime are expected to have lower activity levels and also to present a greater dependency between Rossby numbers and magnetic fields. 
Stars in the saturated regime are in the limit of kinetic to magnetic energy conversion, and show a much flatter relation, although there is still some dependency between Rossby numbers and magnetic fields. 

Most of the stars in our sample have rotational periods roughly between 1 and 10 days. Overall, stars in the saturated regime (blue open circles) follow an approximately constant $<$B$>$ as a function of $P_{\rm rot}$, showing just a modest inclination in the trend. It is clear that our results for this population overlap with results from the literature in this regime. 
For sample stars in the unsaturated regime (open red circles), $<$B$>$ values begin to show a de-saturation signal (at P$_{\rm rot}$ $\sim $7 days), with the slowest rotating stars of our sample presenting a steeper negative relation between rotational periods and magnetic fields than for faster rotators. Although our sample does not include very slowly rotating stars, our results generally overlap with literature values and are in line with the steep relation between magnetic fields and rotational periods in the literature, which extends to P$_{\rm rot}$ $\sim$ 100 days. 
It is interesting to note that despite the fact that our sample and that from R22 both have the same upper $T_{\rm eff}$ limit of 4000 K, the latter sample reaches much greater rotational periods and lower magnetic fields. The possible main reason for this is that R22 results are for field M dwarfs, while our work considers only M dwarf star members of the very young Pleiades open cluster. M dwarfs from the field may have had much more time to lose their magnetic fields and angular momentum when compared to the young stars from our sample.

The dependency between rotation and magnetic fields can be explained by the dynamo mechanism that converts kinetic into magnetic energy. The total magnetic field of the star is limited by its available kinetic energy, and this maximum magnetic field is described as B$_{\rm kin}$. Here we adopt the relation from \citet{reiners2009b} for B$_{\rm kin}$ (in units of Gauss):

\begin{equation}
    B_{\rm kin}=4800 \times \bigg(\frac{ML^{2}}{R^{7}}\bigg)^{1/6}
    \label{eq2}    
\end{equation}

where $L$ is the derived luminosity (in L$_{\odot}$ units), $M$ is the stellar mass (in M$_{\odot}$ units), and $R$ is the radius (in R$_{\odot}$ units) derived using the Stefan-Boltzmann equation, with the adopted T$_{\rm eff}$. 
We derived magnetic fluxes ($\phi_{\rm B}$) by multiplying the obtained average magnetic field by $4 \pi R^{2}$. Masses were determined from isochrones, following the methodology discussed in \citet{wanderley2023}. In summary,
we selected SPOTS isochrones from \citet{Somers2020_spots}  
for 100 Myr and solar metallicity, interpolated isochrones associated with different photospheric spots fractions (f$_{spots}$), and adopted as the mass of the star the point in the interpolated isochrone plane with closest T$_{\rm eff}$ and luminosity of the star. 

We also estimated masses using DARTMOUTH \citep{Dotter2008_dartmouth}, MIST \citep{choi2016_MIST} and BHAC15 \citep{baraffe2015} isochrones, and found that, as expected, the choice of isochrone does not change significantly the B$_{\rm kin}$ results.

\begin{figure*}
\begin{center}
  \includegraphics[angle=0,width=1\linewidth,clip]{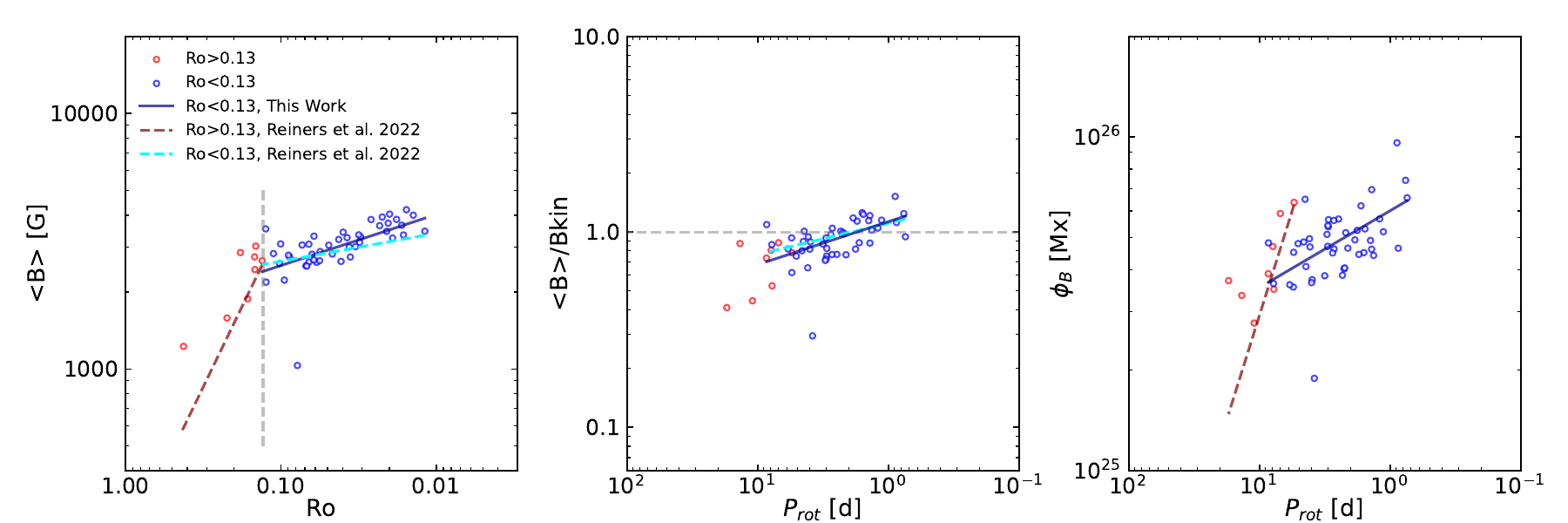}
\caption{Comparison of different magnetic field indicators as a function of Rossby numbers (left panel) and rotational periods (middle and right panels). Stars with Ro$<$0.13 are represented by open blue circles, and stars with Ro$>$0.13 are represented by open red circles.
The left panel presents the distribution of the obtained magnetic fields, the middle panel presents the ratio between the obtained average magnetic fields and the magnetic field limit based on the stellar kinetic energy and the right panel presents the magnetic flux of the stars. The left panel presents a vertical grey line to represent Ro=13. The horizontal grey line in the middle panel represents the point where the magnetic field is at its maximum physical limit, based on kinetic to magnetic energy conversion, where $<$B$>$= B$_{\rm kin}$. We also show relations between rotational and magnetic parameters derived by this work (solid lines) and \citet{reiners2022} (dashed lines), for slow (red lines) and/or rapid rotators (cyan lines).}
\end{center}
\label{triplero}
\end{figure*}

Figure \ref{triplero} presents our results for $<$B$>$ as a function of Rossby numbers (left panel), $<$B$>$/B$_{kin}$ (middle panel), and $\phi_{\rm B}$ (right panel) as a function of rotational periods. Stars with Ro$>$0.13 are represented by open red circles, and stars with Ro$<$0.13 are represented by open blue circles. 
The grey dashed vertical line in the left panel at Ro=0.13 is the threshold that corresponds to the separation between the saturated and the non-saturated regimes.

For stars in the saturated regime in our sample (open blue circles), we derived relations between their mean magnetic fields with Rossby numbers, and between $<$B$>$/B$_{\rm kin}$ and $\phi_{\rm B}$ with rotational periods by using non-linear least squares to fit power-law functions. The obtained relations are found below and these are shown as solid blue lines in Figure \ref{triplero}. 

\begin{equation}
    <B>= 1604 G \times Ro^{-0.20}
    \label{eqb3}
\end{equation}

\begin{equation}
    <B>/B_{\rm kin}= 1.13 \times P_{\rm rot}^{-0.21}
    \label{eqbbkin}
\end{equation}

\begin{equation}
    \phi_{\rm B}= 6.01 \times 10^{25} Mx \times P_{\rm rot}^{-0.23}
    \label{eqphim}
\end{equation}

For comparison, we also show in Figure \ref{triplero} similar relations derived by R22 for their sample of rapid and slow rotators (represented by cyan and red dashed lines, respectively). 

As previously discussed in Figure \ref{litcomp}, the results in the left panel of Figure \ref{triplero} show a clear difference in the trends between the saturated and unsaturated regimes. The relation obtained here for the saturated stars (Equation \ref{eqb3}) is less steep than for the unsaturated ones and similar to the one derived in R22. Given the small number of slow-rotating stars and small range in Rossby number of our sample, we did not compute a best-fit relation for the unsaturated stars but the comparison of our results (open red circles) with the relation in R22 (red dashed line) shows good agreement. 
The results in the middle panel of Figure \ref{triplero} show that overall most of our stars present $<$B$>$/B$_{\rm kin}$ ratios near 1 (represented by the grey horizontal dashed line), indicating that they are near the peak of magnetic energy production based on their available kinetic energy. There is a small trend with rotational periods (or Rossby numbers) showing an inverse correlation between $<$B$>$/B$_{\rm kin}$ and rotational periods. 
This is illustrated by our derived relation for rapid rotators (Equation \ref{eqbbkin}), which is very similar to the one derived by R22 (dashed blue line), presenting a similar slope, and being almost the same as our relation. 
The results in the right panel of Figure \ref{triplero} show an overall similar behavior as found for $<$B$>$ versus Rossby number, with saturated stars presenting a smaller dependency with P$_{\rm rot}$ than non-saturated ones. This is illustrated by the blue solid line showing the relation between magnetic fluxes and rotational periods for the rapid rotator sample (Equation \ref{eqphim}). The well-defined relation obtained here for the saturated regime for the Pleiades M dwarfs is different from the results in R22, who found a large scatter in magnetic fluxes for fast-rotating stars and without a well-defined relation. Such difference in the results may come from the fact that R22 studied field stars with different ages, while our sample has a well-constrained age of $\sim$ 100 million years and, the well-defined trend for the saturated regime may be an indication of a strong connection between magnetic fluxes and stellar ages. The rotational periods and computed Rossby numbers, magnetic fluxes and $<$B$>$/B$_{\rm kin}$ for the stars are presented in Table \ref{compiledata}.

\subsection{Radius Inflation in the Pleiades M dwarfs}

Several works in the literature have compared stellar radii measurements with predicted radii from stellar isochrones that do not include magnetic fields and have found that stars, M dwarf stars in particular, have larger radii than predicted by the models \citep{reiners2012,Jackson2016,Jackson2018,Jackson2019,jeffers2018,Kesseli2018,jaehnig2019,wanderley2023}. Stellar magnetic fields can have the effect of decreasing convective efficiency and/or generating stellar spots, which reduces the stellar photospheric temperature and cause the star to inflate \citep{chabrier2007}. The term radius inflation (R$_{\rm frac}$, fractional radius inflation) refers to the fractional difference between the radius obtained from measurements and isochrone models (R$_{\rm iso}$). 

In this study, we follow the same methodology as discussed in \citet{wanderley2023} to measure radius inflation for our sample of Pleiades M dwarfs, using as baseline MIST, DARTMOUTH, and BHAC15 isochrones with solar-metallicities and ages of 100 Myr. As discussed in Section 4.1, stellar radii were derived from the Stefan-Boltzmann equation, using adopted T$_{\rm eff}$ along with luminosities derived from photometric relations.
The median$\pm$MAD radius inflation obtained in this work is 7.0$\pm$1.5$\%$, 6.5$\pm$1.4$\%$ and 5.4 $\pm$1.3$\%$, respectively, based upon the MIST, DARTMOUTH, and BHAC15 isochrones. A median radius inflation of $\sim$5--7$\%$ found here for the Pleiades M dwarfs is considerably larger than the median radius inflation of $\sim$2$\%$ that was obtained by \citet{wanderley2023} for M-dwarf members of the older (age=625 $\pm$ 50 Myr, \citealt{perryman1998}) Hyades open cluster. 
Greater radius inflation for younger M dwarfs, is generally in line with expectations from gyrochronology that stars from younger clusters should present higher average magnetic fields if compared to older stars with the same masses. 
It should be kept in mind, however, that while the sample of \citet{wanderley2023} included M dwarfs beyond the fully convective threshold, our Pleiades sample is composed only of partially convective M dwarfs, and models including magnetic fields predict that radius inflation can be inhibited in fully convective stars \citep{feiden2015}.

\begin{figure*}
\begin{center}
  \includegraphics[angle=0,width=1\linewidth,clip]{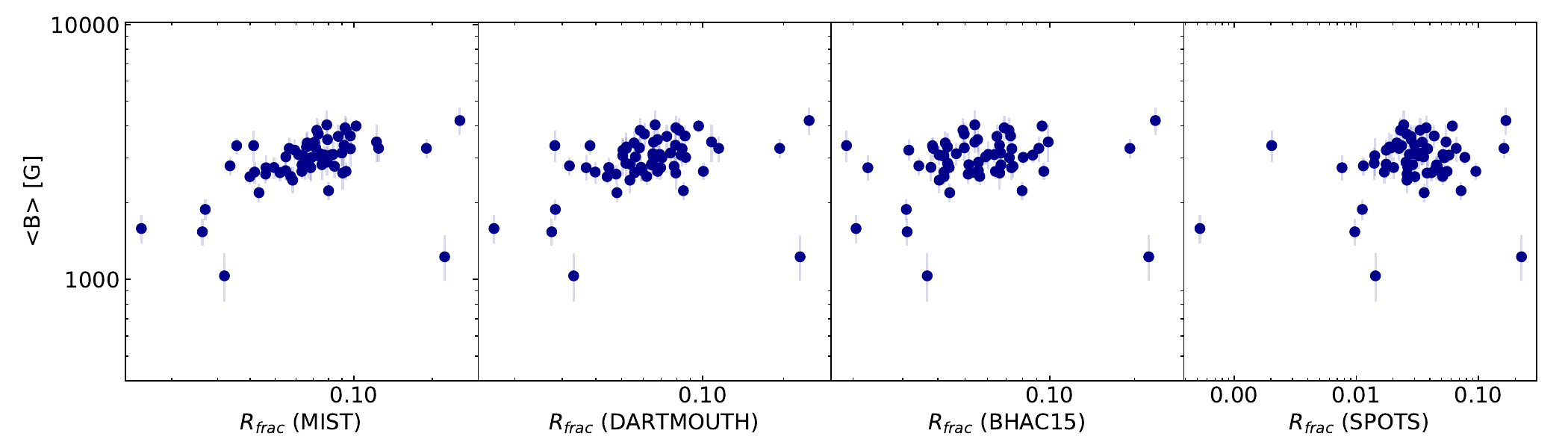}
\caption{Derived average magnetic fields and radius inflation for the sample 62 stars, considering different sets of isochrones. From left to right radius inflation is based on respectively MIST, DARTMOUTH, BHAC15, and SPOTS isochrones.}
\end{center}
\label{radinf}
\end{figure*}

In addition, in this study, we also computed radius inflation using as a baseline SPOTS isochrone models from \citet{Somers2020_spots}. These models consider that stellar spots change the internal structure of the star by suppressing convection, reducing the photospheric temperature, and as a consequence inflating the star. As previously, to derive the fractional radius inflation (R$_{\rm frac}$) and photospheric spot fractions (f$_{\rm spots}$), we adopted SPOTS isochrones for solar metallicity, and an age of 100 Myr and the same methodology to measure radius inflation presented in \citet{wanderley2023} (We refer to this previous study for details). 

Using SPOTS models, we find a median radius inflation for our sample of 3.0$\pm$1.2$\%$, which is smaller when compared to that obtained with either the MIST, DARTMOUTH or BHAC15 isochrones. Less radius inflation is expected when using \citet{Somers2020_spots} isochrones as a baseline due to the decrease in luminosity for more spotted models. The median radius inflation found here for the Pleiades M dwarfs is larger than the median radius inflation of 1.0$\pm$0.5$\%$ reported by \citet{wanderley2023} for the Hyades. 

\begin{figure}
\begin{center}
  \includegraphics[angle=0,width=1\linewidth,clip]{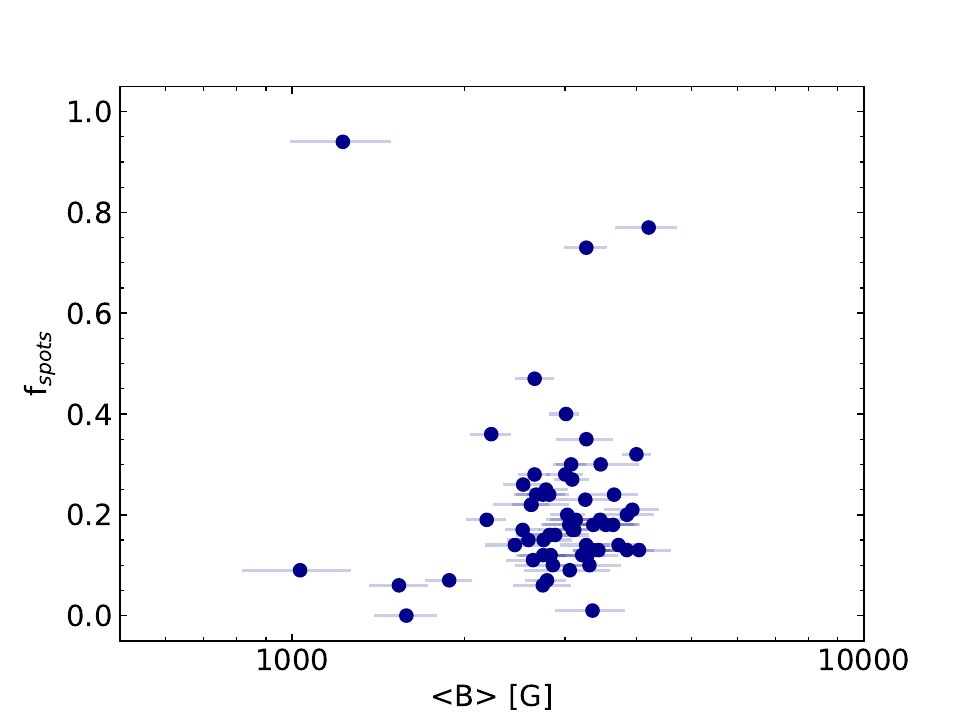}
\caption{Fractional stellar photospheric spots coverage from SPOTS models \citep{Somers2020_spots}, as a function of our derived average magnetic fields.}
\end{center}
\label{spots}
\end{figure}

Figure \ref{radinf} presents our results for the mean magnetic fields and derived radius inflation (R$_{\rm frac}$) for the sample M dwarfs. Panels from left to right show the radius inflation for the MIST, DARTMOUTH, BHAC15, and SPOTS isochrones. The main result from all panels is that overall Pleiades M dwarfs having higher mean magnetic fields have more inflated radii.
The work of \citet{feiden2015} applied modifications in the DARTMOUTH models to include the effects of magnetic fields and found a positive correlation between magnetic field strength and radius inflation when considering either rotational or turbulent dynamos. This finding agrees with our results, as for all studied isochrones, stars with higher magnetic fields tend to be more inflated. There is, however, a tendency for $<$B$>$ to have a weaker dependency with radius inflation, in particular for the SPOTS models with R$_{\rm frac}$ between 0.02 and 0.10.
Figure \ref{spots} presents the derived photospheric stellar spot fractions as a function of average magnetic fields. For the majority of the Pleiades M dwarfs, there is a positive correlation between stellar spot fractions and magnetic field, although with a small number of outliers. 
Finally, we note that there is one clear outlier star in Figures \ref{radinf} and \ref{spots}, having high radius inflation and low $<$B$>$. This star has the lowest K$_s$ mag from our sample (K$_s$ mag = 10.42; \citet{skrustkie2006}). Its position the J-K$_s$ vs K$_s$ diagram (see Figure \ref{membership}) might hint that it is not a member of the Pleiades cluster. However, this star was found in \citet{CANTATGAUDIN_2020} to have a 100$\%$ probability of being a member of the Pleiades.
Note that our f$_{\rm spots}$ may progressively increase for $<$B$>$ larger than roughly 2500 G. Although,
\citet{cao2022} previously found constant f$_{\rm spots}$ at low Rossby number (Ro $<$ 0.21) in their analysis of Pleiades dwarfs with their values of f$_{\rm spots}$ modeled from a two-component surface defined by a star-spot filling factor and a star-spot temperature contrast.

\subsection{Magnetic Fields and Activity}

Stellar activity is a term that represents the stellar variability phenomenon which is mainly a consequence of strong magnetic fields. Stellar variability can be caused by stellar spots, which are temporary cooler regions in the stellar photosphere where the strong magnetic fields suppress convection. As the star rotates these cold spots can reduce the observed stellar flux, and result in periodic changes in the magnitude of the star. 
Another important mechanism that can generate stellar variability is high-energy non-thermal emission. Strong magnetic fields are responsible for heating the coronae and chromosphere of active stars, which results in considerably larger amounts of high-energy radiation emission than expected from their black-body profiles. Heating of the coronae results in excess in the X-ray stellar output that can be measured as the ratio between the X-ray and bolometric luminosity, while heating of the chromosphere is often studied from magnetic sensitive emission lines, such as H$\alpha$, H$\beta$, and Ca II H and K lines. Since coronal and chromospheric non-thermal emissions are produced by the effect of stellar magnetic fields, they are also variable, and change according to the stellar magnetic cycle.

\begin{figure}
\begin{center}
  \includegraphics[angle=0,width=1\linewidth,clip]{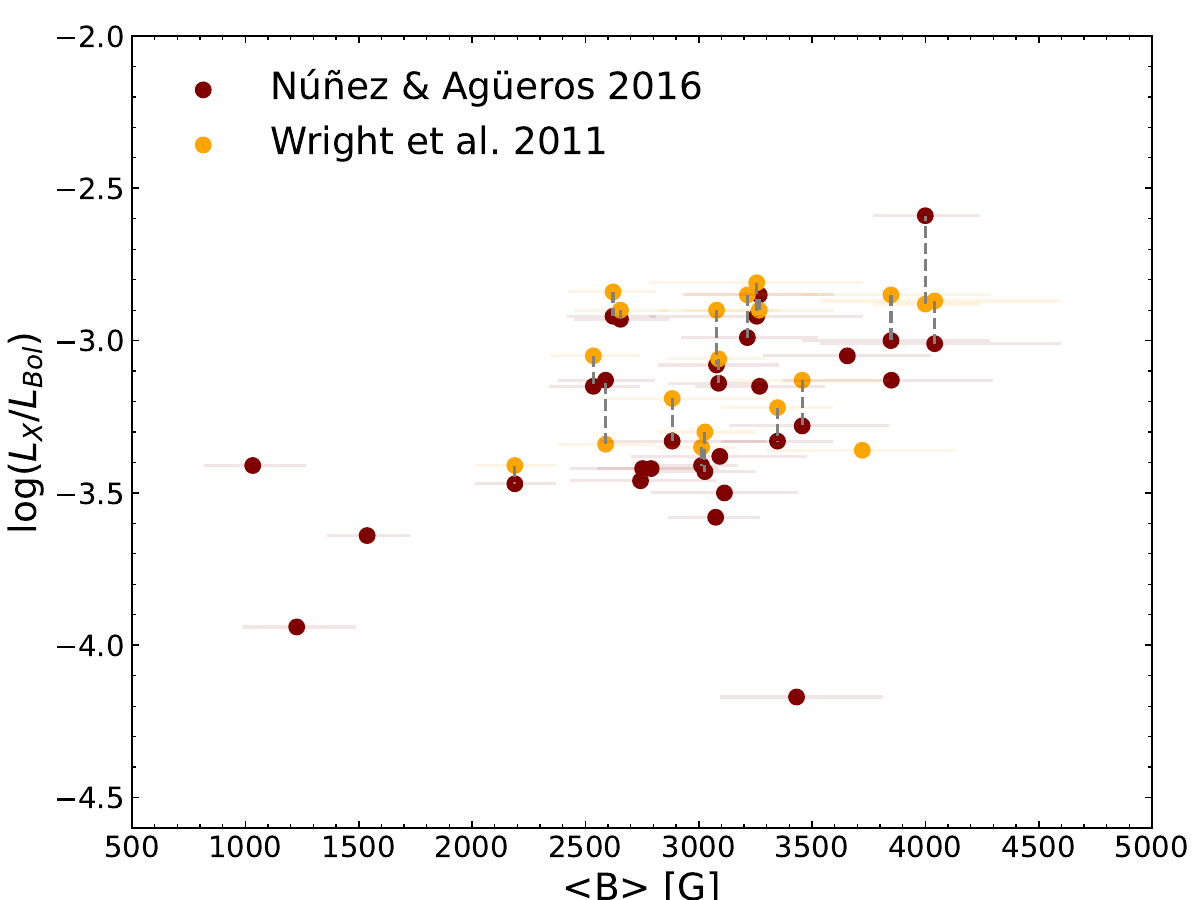}
\caption{Derived average magnetic fields as a function of the ratios between X-ray to bolometric luminosities for stars of our sample in common with the study of \citet{wright2011} (orange circles), and \citet{nunesagueros2016} (maroon circles). The same stars are connected by dashed lines.}
\end{center}
\label{corona} 
\end{figure}

To study the relation between activity and magnetic fields in the Pleiades M dwarfs, we cross-matched our sample with the targets in the X-ray studies by \citet{nunesagueros2016} and \citet{wright2011}, and found 31 stars in common with the first study (not considering 17 stars having only upper limit measurements), and 19 stars in common with the latter study; the X-ray to bolometric luminosity ratios in those works are X-ray flux measurements from Einstein and ROSAT observations \citep{micela1990,stauffer1994,micela1996,micela1999,stelzer2000}. 
The results are summarized in the top panel of Figure \ref{corona} where we show the logarithm of the X-ray to bolometric luminosity ratios versus our derived magnetic fields. 
There is a clear correlation between the two variables, with a Pearson correlation coefficient of 0.46 for the sample in \citet{nunesagueros2016} (shown as maroon circles) and 0.35 for the sample in \citet{wright2011} (shown as orange circles), indicating that stars with stronger magnetic fields tend to present greater X-ray to bolometric luminosity ratios.

\begin{figure}
\begin{center}
  \includegraphics[angle=0,width=1\linewidth,clip]{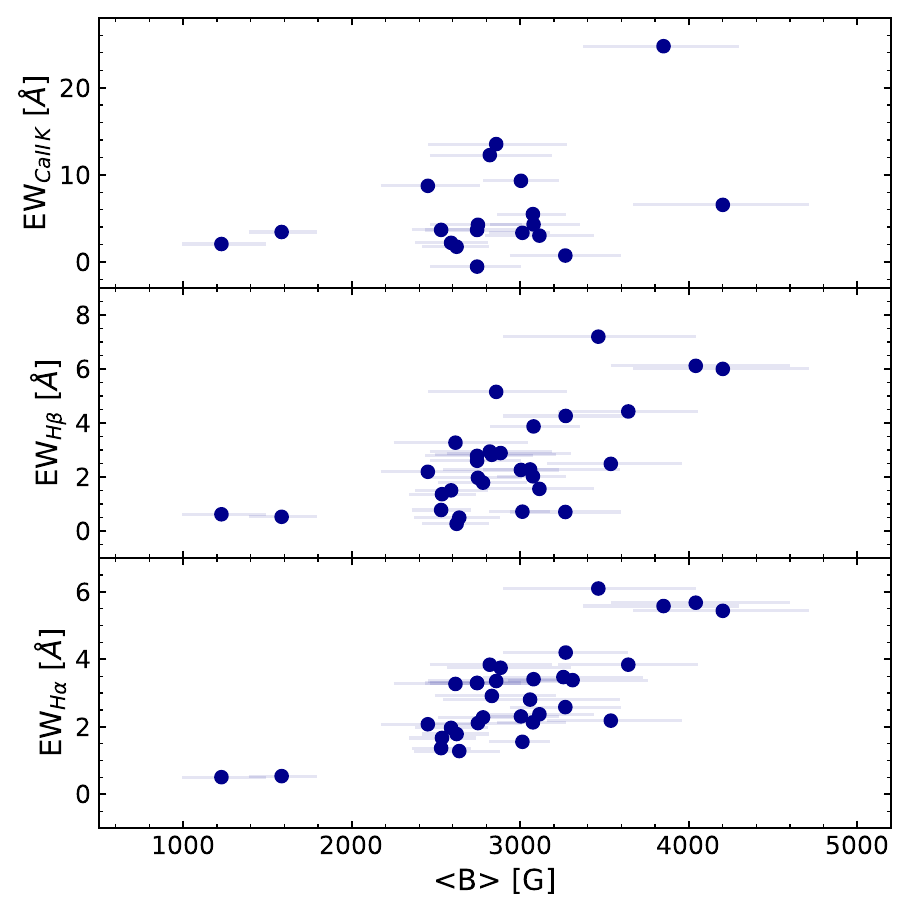}
\caption{Derived average magnetic fields as a function of equivalent widths of magnetic sensitive chromospheric emission lines for stars of our sample that are in common with the work of \citet{fang2018}. The panels present from top to bottom: Ca II K, H$\beta$ and H$\alpha$ emission lines.}
\end{center}
\label{chromo}
\end{figure}

Concerning chromospheric activity the work of \citet{fang2018} measured equivalent widths for H$\alpha$, H$\beta$, and Ca II K emission lines for stars in open clusters using LAMOST DR3 spectra \citep{cui2012}. We cross-matched our Pleiades sample with their database and found 33 stars in common with measured H$\alpha$, 31 with measured H$\beta$, and 19 with measured Ca II K. The three panels in Figure \ref{chromo} present our derived average magnetic field measurements versus the total equivalent width for the Ca II K line (top panel), H$\beta$ (middle panel), and H$\alpha$ (bottom panel), all equivalent widths are in units of \r{A}. Similarly to what was found for the X-ray to bolometric ratios, there is a correlation between the mean magnetic fields for the Pleiades M dwarfs and the equivalent widths of the emission lines for all three lines, although the results for the Ca II K line are less clear. 

\begin{figure}
\begin{center}
  \includegraphics[angle=0,width=1\linewidth,clip]{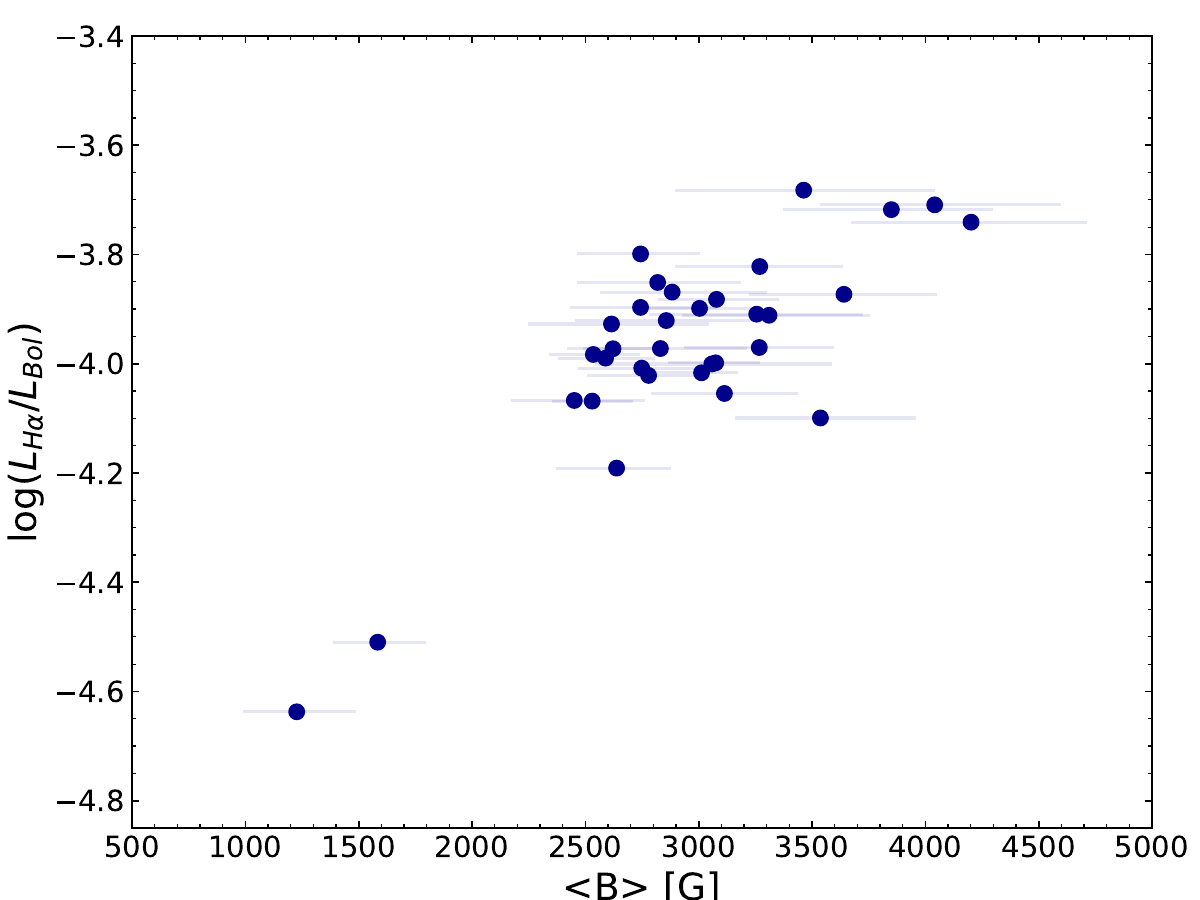}
\caption{Derived average magnetic fields as a function of H$\alpha$ to bolometric luminosities. We employed the methodology described in \citet{stassun2012} to convert H$\alpha$ equivalent widths (taken from \citet{fang2018}) into L$_{H \alpha}$/L$_{\rm bol}$.}
\end{center}
\label{halpha}
\end{figure}

Figure \ref{halpha} shows the H$\alpha$ to bolometric luminosity ratios as a function of mean magnetic field for stars in our sample having H$\alpha$ emission strength measurements (EW$_{H \alpha}$) reported in Table \ref{compiledata}. 
To convert H$\alpha$ equivalent widths into L$_{H \alpha}$/L$_{\rm bol}$ we employed the methodology described in \citet{stassun2012}. Similarly to what is seen for the relation between EW$_{H \alpha}$ and $<$B$>$ (bottom panel of Figure \ref{chromo}), there is a clear correlation between H$\alpha$ to bolometric luminosity ratios and the derived magnetic fields, with a Pearson correlation coefficient of 0.78.

Overall, it is reassuring that in this study we find that the more magnetic stars in our sample tend also to be more active, and this is based on our magnetic field measurements and independent results from activity indicators from other works in the literature. 
The X-ray and H$\alpha$ to bolometric luminosity ratios as well as the equivalent widths for H$\alpha$, H$\beta$ and Ca II K of the stars are presented in Table \ref{compiledata}.

\section{Conclusions}

We used the SYNMAST code \cite{kochukhov2010_synmast}, along with a Markov Chain Monte Carlo (MCMC) methodology, to compute synthetic spectra and analyze magnetically sensitive Fe I lines to derive average magnetic fields for 62 M dwarf members of the young (age = 112 $\pm$ 5 Myr; \citealt{dahm2015_age4}) and nearby Pleiades open cluster. This analysis is based on the SDSS IV APOGEE spectra \citep{majewski2017_apogee,apogeedr17_2022}, the APOGEE line list \citep{smith2021}, effective Landé-g factors from VALD \citep{piskunov1995,kupka1999}, and model atmospheres from the MARCS grid \citep{gustafsson2008_marcs}. 

A search was carried out to find the best Fe I lines in the APOGEE spectral region that were sensitive to Zeeman broadening and could be used as diagnostic lines for measuring mean magnetic fields in M dwarfs, with four Fe I lines identified and selected as the best indicators: $\lambda$15207.526 \r{A}, $\lambda$15294.56 \r{A}, $\lambda$15621.654 \r{A}, and $\lambda$15631.948 \r{A}. The derived mean magnetic fields for the studied Pleiades M dwarf stars range from $\sim$ 1.0 to $\sim$ 4.2 kG, with a median$\pm$MAD of 3.0$\pm$0.3 kG, not reaching the lowest magnetic field and rotation levels reported in other studies that explored field stars of similar masses, which is probably explained by the young age of the cluster.

The derived mean magnetic field measurements in the Pleiades M dwarfs were used to study correlations with Rossby number (Ro) and stellar rotation. 
The Rossby number is given by the ratio between the rotational period and the convective turnover time, being an important indicator of stellar activity.  We find a clear trend that more magnetic stars have, on average, higher projected rotational velocities, v$\sin{i}$. 
The Rossby number was used to separate our sample into rapid (Ro$<$0.13) and slow rotators (Ro$>$0.13). Overall, our results for $<$B$>$ versus P$_{\rm rot}$ and B versus Rossby number overlap with results from the literature for field stars, and indicate that the population of stars with Ro$<$0.13 exhibit a steeper relation between magnetic field and rotational period, or Rossby number, when compared to stars with Ro$>$0.13.  However, even for Ro$>$0.13, there remains a shallow trend between Rossby number and magnetic field, which is given by: $<$B$>$ = $1604 \times $Ro$^{-0.20}$.

For this sample of Pleiades M dwarfs, we also investigated the ratio between mean magnetic fields and the maximum magnetic field limit ($<$B$>$/B$_{\rm kin}$) that is reached based on the kinetic-to-magnetic energy conversion.  It is found that most of the studied Pleiades M dwarfs are at the limit of kinetic to magnetic energy conversion, or, are in the saturated regime, having $<$B$>$/B$_{\rm kin} \approx$1. Unlike previous results in the literature for field stars in the saturated regime, the computed stellar magnetic fluxes $\phi_{\rm B}$ as a function of P${\rm rot}$ for the Pleiades M dwarfs show similar trends obtained for $<$B$>$ versus P${\rm rot}$.  

Another important effect of magnetic fields is to inflate the stellar radii of cool dwarfs. Radius inflation corresponds to the fractional difference (R$_{\rm frac}$) between the radius obtained from measurements and predictions from isochrone models.
In this study, we derived the radii for the studied M dwarfs and used MIST, DARTMOUTH, BHAC15, and SPOTS isochrones as baselines to infer their radius inflation. We obtain a median$\pm$MAD radius inflation for our sample of respectively 7.0$\pm$1.5$\%$, 6.5$\pm$1.4$\%$, 5.4 $\pm$1.3$\%$ and 3.0 $\pm$1.2$\%$, being more inflated than M dwarfs in the older Hyades open cluster \citet{wanderley2023}. For the Pleiades, it is noticeable that for SPOTS isochrones there is less radius inflation when compared to non-spotted isochrones, as expected. In addition, our results indicate that more magnetic stars have more inflated radii, showing a correlation between R$_{\rm frac}$ and $<$B$>$. For SPOTS models in particular, there is, however, a tendency for $<$B$>$ to exhibit a weaker dependency with radius inflation for R$_{\rm frac}$ between 0.02 and 0.10.
For the majority of the Pleiades M dwarfs in our sample, there is a positive correlation between stellar spot fraction and magnetic field, although with a small number of outliers. 

To study the relation between chromospheric stellar activity and magnetic fields in the Pleiades M dwarfs, we compared our derived mean magnetic fields with high-energy non-thermal emission indicators, such as equivalent width measurements of the lines of H$\alpha$, H$\beta$, and Ca II K, as well as H$\alpha$ to bolometric luminosity ratios. For all of these, we find a positive correlation between chromospheric activity and magnetic fields.  A positive correlation was also obtained between the mean magnetic field and the ratio between X-ray to bolometric luminosity, which is an important indicator of coronal activity.  Overall, it is reassuring that the more magnetic stars in this study also tend to be more active, and this is based on our magnetic field measurements from APOGEE spectra, which is independent of the results for activity indicators obtained from the literature.  This study opens a new window into using the APOGEE survey to investigate magnetic fields in cool stars.

\acknowledgments

We thank the referee for the comments that improved the paper.
F.W. acknowledges support from fellowship by Coordena\c c\~ao de Ensino Superior - CAPES. 
K.C. and V.S. acknowledge that their work is supported,
in part, by the National Science Foundation through NSF grant No. AST-2009507.
O.K. acknowledges support by the Swedish Research Council (grant agreements no. 2019-03548 and 2023-03667), the Swedish National Space Agency, and the Royal Swedish Academy of Sciences.
D.S. thanks the National Council for Scientific and Technological Development – CNPq.

Funding for the Sloan Digital Sky Survey IV has been provided by the Alfred P. Sloan Foundation, the U.S. Department of Energy Office of Science, and the Participating Institutions. SDSS-IV acknowledges support and resources from the Center for High-Performance Computing at the University of Utah. The SDSS web site is www.sdss.org.
SDSS-IV is managed by the Astrophysical Research consortium for the Participating Institutions of the SDSS Collaboration including the Brazilian Participation Group, the Carnegie Institution for Science, Carnegie Mellon University, the Chilean Participation Group, the French Participation Group, Harvard-Smithsonian Center for Astrophysics, Instituto de Astrof\'isica de Canarias, The Johns Hopkins University, 
Kavli Institute for the Physics and Mathematics of the Universe (IPMU) /  University of Tokyo, Lawrence Berkeley National Laboratory, Leibniz Institut f\"ur Astrophysik Potsdam (AIP),  Max-Planck-Institut f\"ur Astronomie (MPIA Heidelberg), Max-Planck-Institut f\"ur Astrophysik (MPA Garching), Max-Planck-Institut f\"ur Extraterrestrische Physik (MPE), National Astronomical Observatory of China, New Mexico State University, New York University, University of Notre Dame, Observat\'orio Nacional / MCTI, The Ohio State University, Pennsylvania State University, Shanghai Astronomical Observatory, United Kingdom Participation Group,
Universidad Nacional Aut\'onoma de M\'exico, University of Arizona, University of Colorado Boulder, University of Oxford, University of Portsmouth, University of Utah, University of Virginia, University of Washington, University of Wisconsin, Vanderbilt University, and Yale University.

\facility {Sloan}

\software{Matplotlib (\citealt{Hunter2007_matplotlib}), Numpy (\citealt{harris2020_numpy}), Synmast (\citealt{kochukhov2010_synmast}), Turbospectrum (\citealt{plez2012_turbospectrum})}

\startlongtable
\begin{deluxetable*}{lccccccccc}
\small\addtolength{\tabcolsep}{-1pt}
\tablenum{1}
\label{filfactor}
\tabletypesize{\scriptsize}
\tablecaption{Filling Factors}
\tablewidth{0pt}
\tablehead{
\colhead{APOGEE ID} &
\colhead{f$_{0}$} &
\colhead{f$_{2}$} &
\colhead{f$_{4}$} &
\colhead{f$_{6}$} &
\colhead{f$_{8}$} &
\colhead{f$_{10}$} &
\colhead{f$_{12}$} &
\colhead{$<$B$>$} \\
\colhead{...} &
\colhead{...} &
\colhead{...} &
\colhead{...} &
\colhead{...} &
\colhead{...} & 
\colhead{...} & 
\colhead{...} & 
\colhead{G} & 
}
\startdata
2M03420291+2355538 & 0.218 $\pm$0.087 & 0.274 $\pm$0.152 & 0.147 $\pm$0.134 & 0.034 $\pm$0.153 & 0.246 $\pm$0.22 & 0.037 $\pm$0.049 & 0.044 $\pm$0.098 & 4201 $^{511}_{529}$ \\
2M03481801+2353294 & 0.784 $\pm$0.058 & 0.113 $\pm$0.064 & 0.037 $\pm$0.03 & 0.011 $\pm$0.011 & 0.013 $\pm$0.01 & 0.005 $\pm$0.025 & 0.036 $\pm$0.016 & 1032 $^{236}_{215}$ \\
\enddata
\end{deluxetable*}

\startlongtable
\begin{deluxetable*}{lcccccccccccccc}
\small\addtolength{\tabcolsep}{-2.8pt}
\tablenum{2}
\label{compiledata}
\tabletypesize{\scriptsize}
\tablecaption{Stellar Data}
\tablewidth{0pt}
\tablehead{
\colhead{APOGEE ID} &
\colhead{SNR} &
\colhead{v$\sin{i}$} &
\colhead{T$_{\rm eff}$} &
\colhead{P$_{\rm rot}$} &
\colhead{Ro} &
\colhead{$<$B$>$} &
\colhead{$<$B$>$/B$_{\rm kin}$} &
\colhead{$\phi_{\rm B}$} &
\colhead{$\log{(L_{\rm X}/L_{\rm bol})}$} &
\colhead{$\log{(L_{H \alpha}/L_{\rm bol})}$} &
\colhead{EW$_{H\alpha}$} &
\colhead{EW$_{H\beta}$} &
\colhead{EW$_{\rm CaIIK}$} & \\
\colhead{...} &
\colhead{...} &
\colhead{km s$^{-1}$} &
\colhead{K} &
\colhead{day} &
\colhead{...} &
\colhead{G} &
\colhead{...} &
\colhead{$10^{25}$ Mx} &
\colhead{...} &
\colhead{...} &
\colhead{\r{A}} &
\colhead{\r{A}} &
\colhead{\r{A}} &
}
\startdata
2M03420291+2355538 & 138 & 17.7 & 3424 & 0.89 & 0.02 & 4202$^{+511}_{-530}$ & 1.52 & 9.6 & ...$^{a}$/-2.92$^{b}$  & 5.43 & -3.74 & 6 & 6.55 \\
2M03424379+2532064 & 103 & 7.2 & 3443 & ... & ... & 3362$^{+669}_{-615}$ & 1.06 & 3.81 & ...$^{a}$/...$^{b}$  & ... & ... & ... & ... \\
2M03515114+2317414 & 112 & 13.4 & 3452 & 0.87 & 0.01 & 3463$^{+581}_{-567}$ & 1.12 & 4.64 & ...$^{a}$/...$^{b}$  & 6.1 & -3.68 & 7.2 & ... \\
2M03422864+2501004 & 131 & 15.2 & 3462 & 1.39 & 0.02 & 3850$^{+449}_{-478}$ & 1.21 & 4.61 & ...$^{a}$/-3.13$^{b}$  & 5.58 & -3.72 & 9.46 & 24.76 \\
... & ... & ... & ... & ... & ... & ... & ... & ... & ... & ... & ... & ... & ... \\
\enddata
\begin{tablenotes}
\item \normalsize The complete Table is available in electronic format.
\item \small ${}^{a}_{}$\citet{wright2011}
\item \small ${}^{b}_{}$\citet{nunesagueros2016}
\end{tablenotes}
\end{deluxetable*}

\startlongtable
\begin{deluxetable*}{lcccccc}
\tablenum{3}
\label{linelist}
\tabletypesize{\scriptsize}
\tablecaption{Diagnostic lines}
\tablewidth{0pt}
\tablehead{
\colhead{$\lambda$} &
\colhead{$\chi_{\rm exc}$} &
\colhead{$\log{\rm gf}$} &
\colhead{Effective Landé-g} &
\colhead{Lower level term designation} &
\colhead{Upper level term designation} \\
\colhead{$\mathring{A}$} &
\colhead{eV} &
\colhead{...} &
\colhead{...} &
\colhead{...} &
\colhead{...} & 
}
\startdata
15207.526 & 5.3852 & 0.067 & 1.532 & LS 3d6.(5D).4s\ (6D).5s e7D	& LS 3d6.(5D).4s\ (6D).5p n7D* \\
15294.560 & 5.3085 & 0.523 & 1.590 & LS 3d6.(5D).4s\ (6D).5s e7D	& LS 3d6.(5D).4s\ (6D).5p n7D* \\
15621.654 & 5.5392 & 0.280 & 1.494 & LS 3d6.(5D).4s\ (6D).5s e5D	& LS 3d6.(5D).4s\ (6D).5p t5D* \\
15631.948 & 5.3516 & -0.032 & 1.655 & LS 3d6.(5D).4s\ (6D).5s e7D	& LS 3d6.(5D).4s\ (6D).5p n7D* \\
\enddata
\end{deluxetable*}

\bibliographystyle{yahapj}
\bibliography{references.bib}

\begin{thebibliography}{}
\providecommand\natexlab[1]{#1}
\providecommand\JournalTitle[1]{#1}

\bibitem[{{Abdurro'uf} {et~al.}(2022){Abdurro'uf}, {Accetta}, {Aerts}, {Silva Aguirre}, {Ahumada}, {Ajgaonkar}, {Filiz Ak}, {Alam}, {Allende Prieto}, {Almeida}, {Anders}, {Anderson}, {Andrews}, {Anguiano}, {Aquino-Ort{\'\i}z}, {Arag{\'o}n-Salamanca}, {Argudo-Fern{\'a}ndez}, {Ata}, {Aubert}, {Avila-Reese}, {Badenes}, {Barb{\'a}}, {Barger}, {Barrera-Ballesteros}, {Beaton}, {Beers}, {Belfiore}, {Bender}, {Bernardi}, {Bershady}, {Beutler}, {Bidin}, {Bird}, {Bizyaev}, {Blanc}, {Blanton}, {Boardman}, {Bolton}, {Boquien}, {Borissova}, {Bovy}, {Brandt}, {Brown}, {Brownstein}, {Brusa}, {Buchner}, {Bundy}, {Burchett}, {Bureau}, {Burgasser}, {Cabang}, {Campbell}, {Cappellari}, {Carlberg}, {Wanderley}, {Carrera}, {Cash}, {Chen}, {Chen}, {Cherinka}, {Chiappini}, {Choi}, {Chojnowski}, {Chung}, {Clerc}, {Cohen}, {Comerford}, {Comparat}, {da Costa}, {Covey}, {Crane}, {Cruz-Gonzalez}, {Culhane}, {Cunha}, {Dai}, {Damke}, {Darling}, {Davidson}, {Davies}, {Dawson}, {De Lee}, {Diamond-Stanic}, {Cano-D{\'\i}az}, {S{\'a}nchez},
  {Donor}, {Duckworth}, {Dwelly}, {Eisenstein}, {Elsworth}, {Emsellem}, {Eracleous}, {Escoffier}, {Fan}, {Farr}, {Feng}, {Fern{\'a}ndez-Trincado}, {Feuillet}, {Filipp}, {Fillingham}, {Frinchaboy}, {Fromenteau}, {Galbany}, {Garc{\'\i}a}, {Garc{\'\i}a-Hern{\'a}ndez}, {Ge}, {Geisler}, {Gelfand}, {G{\'e}ron}, {Gibson}, {Goddy}, {Godoy-Rivera}, {Grabowski}, {Green}, {Greener}, {Grier}, {Griffith}, {Guo}, {Guy}, {Hadjara}, {Harding}, {Hasselquist}, {Hayes}, {Hearty}, {Hern{\'a}ndez}, {Hill}, {Hogg}, {Holtzman}, {Horta}, {Hsieh}, {Hsu}, {Hsu}, {Huber}, {Huertas-Company}, {Hutchinson}, {Hwang}, {Ibarra-Medel}, {Chitham}, {Ilha}, {Imig}, {Jaekle}, {Jayasinghe}, {Ji}, {Johnson}, {Jones}, {J{\"o}nsson}, {Katkov}, {Khalatyan}, {Kinemuchi}, {Kisku}, {Knapen}, {Kneib}, {Kollmeier}, {Kong}, {Kounkel}, {Kreckel}, {Krishnarao}, {Lacerna}, {Lane}, {Langgin}, {Lavender}, {Law}, {Lazarz}, {Leung}, {Leung}, {Lewis}, {Li}, {Li}, {Lian}, {Liang}, {Lin}, {Lin}, {Lin}, {Lintott}, {Long}, {Longa-Pe{\~n}a}, {L{\'o}pez-Cob{\'a}}, {Lu},
  {Lundgren}, {Luo}, {Mackereth}, {de la Macorra}, {Mahadevan}, {Majewski}, {Manchado}, {Mandeville}, {Maraston}, {Margalef-Bentabol}, {Masseron}, {Masters}, {Mathur}, {McDermid}, {Mckay}, {Merloni}, {Merrifield}, {Meszaros}, {Miglio}, {Di Mille}, {Minniti}, {Minsley}, {Monachesi}, {Moon}, {Mosser}, {Mulchaey}, {Muna}, {Mu{\~n}oz}, {Myers}, {Myers}, {Nadathur}, {Nair}, {Nandra}, {Neumann}, {Newman}, {Nidever}, {Nikakhtar}, {Nitschelm}, {O'Connell}, {Garma-Oehmichen}, {Luan Souza de Oliveira}, {Olney}, {Oravetz}, {Ortigoza-Urdaneta}, {Osorio}, {Otter}, {Pace}, {Padilla}, {Pan}, {Pan}, {Parikh}, {Parker}, {Peirani}, {Pe{\~n}a Ram{\'\i}rez}, {Penny}, {Percival}, {Perez-Fournon}, {Pinsonneault}, {Poidevin}, {Poovelil}, {Price-Whelan}, {B{\'a}rbara de Andrade Queiroz}, {Raddick}, {Ray}, {Rembold}, {Riddle}, {Riffel}, {Riffel}, {Rix}, {Robin}, {Rodr{\'\i}guez-Puebla}, {Roman-Lopes}, {Rom{\'a}n-Z{\'u}{\~n}iga}, {Rose}, {Ross}, {Rossi}, {Rubin}, {Salvato}, {S{\'a}nchez}, {S{\'a}nchez-Gallego}, {Sanderson}, {Santana
  Rojas}, {Sarceno}, {Sarmiento}, {Sayres}, {Sazonova}, {Schaefer}, {Schiavon}, {Schlegel}, {Schneider}, {Schultheis}, {Schwope}, {Serenelli}, {Serna}, {Shao}, {Shapiro}, {Sharma}, {Shen}, {Shetrone}, {Shu}, {Simon}, {Skrutskie}, {Smethurst}, {Smith}, {Sobeck}, {Spoo}, {Sprague}, {Stark}, {Stassun}, {Steinmetz}, {Stello}, {Stone-Martinez}, {Storchi-Bergmann}, {Stringfellow}, {Stutz}, {Su}, {Taghizadeh-Popp}, {Talbot}, {Tayar}, {Telles}, {Teske}, {Thakar}, {Theissen}, {Tkachenko}, {Thomas}, {Tojeiro}, {Hernandez Toledo}, {Troup}, {Trump}, {Trussler}, {Turner}, {Tuttle}, {Unda-Sanzana}, {V{\'a}zquez-Mata}, {Valentini}, {Valenzuela}, {Vargas-Gonz{\'a}lez}, {Vargas-Maga{\~n}a}, {Alfaro}, {Villanova}, {Vincenzo}, {Wake}, {Warfield}, {Washington}, {Weaver}, {Weijmans}, {Weinberg}, {Weiss}, {Westfall}, {Wild}, {Wilde}, {Wilson}, {Wilson}, {Wilson}, {Wolf}, {Wood-Vasey}, {Yan}, {Zamora}, {Zasowski}, {Zhang}, {Zhao}, {Zheng}, {Zheng}, \& {Zhu}}]{apogeedr17_2022}
{Abdurro'uf}, {Accetta}, K., {Aerts}, C., {et~al.} 2022, \href{http://dx.doi.org/10.3847/1538-4365/ac4414}{\JournalTitle{\apjs}, 259, 35}

\bibitem[{{Afram} {et~al.}(2009){Afram}, {Reiners}, \& {Berdyugina}}]{afram2009_olegreview9}
{Afram}, N., {Reiners}, A., \& {Berdyugina}, S.~V. 2009, in Astronomical Society of the Pacific Conference Series, Vol. 405, Solar Polarization 5: In Honor of Jan Stenflo, ed. S.~V. {Berdyugina}, K.~N. {Nagendra}, \& R.~{Ramelli}, 527

\bibitem[{{Astudillo-Defru} {et~al.}(2017){Astudillo-Defru}, {Delfosse}, {Bonfils}, {Forveille}, {Lovis}, \& {Rameau}}]{astudillo2017}
{Astudillo-Defru}, N., {Delfosse}, X., {Bonfils}, X., {et~al.} 2017, \href{http://dx.doi.org/10.1051/0004-6361/201527078}{\JournalTitle{\aap}, 600, A13}

\bibitem[{{Bailer-Jones} {et~al.}(2021){Bailer-Jones}, {Rybizki}, {Fouesneau}, {Demleitner}, \& {Andrae}}]{bailerjones2021}
{Bailer-Jones}, C.~A.~L., {Rybizki}, J., {Fouesneau}, M., {Demleitner}, M., \& {Andrae}, R. 2021, \href{http://dx.doi.org/10.3847/1538-3881/abd806}{\JournalTitle{\aj}, 161, 147}

\bibitem[{{Bakos} {et~al.}(2004){Bakos}, {Noyes}, {Kov{\'a}cs}, {Stanek}, {Sasselov}, \& {Domsa}}]{bakos2004}
{Bakos}, G., {Noyes}, R.~W., {Kov{\'a}cs}, G., {et~al.} 2004, \href{http://dx.doi.org/10.1086/382735}{\JournalTitle{\pasp}, 116, 266}

\bibitem[{{Baraffe} {et~al.}(2015){Baraffe}, {Homeier}, {Allard}, \& {Chabrier}}]{baraffe2015}
{Baraffe}, I., {Homeier}, D., {Allard}, F., \& {Chabrier}, G. 2015, \href{http://dx.doi.org/10.1051/0004-6361/201425481}{\JournalTitle{\aap}, 577, A42}

\bibitem[{{Barnes}(2003)}]{barnes2003}
{Barnes}, S.~A. 2003, \href{http://dx.doi.org/10.1086/367639}{\JournalTitle{\apj}, 586, 464}

\bibitem[{{Basri} \& {Marcy}(1994)}]{basri1994}
{Basri}, G., \& {Marcy}, G.~W. 1994, \href{http://dx.doi.org/10.1086/174535}{\JournalTitle{\apj}, 431, 844}

\bibitem[{{Basri} {et~al.}(1992){Basri}, {Marcy}, \& {Valenti}}]{basri1992}
{Basri}, G., {Marcy}, G.~W., \& {Valenti}, J.~A. 1992, \href{http://dx.doi.org/10.1086/171312}{\JournalTitle{\apj}, 390, 622}

\bibitem[{{Belokurov} {et~al.}(2020){Belokurov}, {Penoyre}, {Oh}, {Iorio}, {Hodgkin}, {Evans}, {Everall}, {Koposov}, {Tout}, {Izzard}, {Clarke}, \& {Brown}}]{belokurov2020}
{Belokurov}, V., {Penoyre}, Z., {Oh}, S., {et~al.} 2020, \href{http://dx.doi.org/10.1093/mnras/staa1522}{\JournalTitle{\mnras}, 496, 1922}

\bibitem[{{Blanton} {et~al.}(2017){Blanton}, {Bershady}, {Abolfathi}, {Albareti}, {Allende Prieto}, {Almeida}, {Alonso-Garc{\'\i}a}, {Anders}, {Anderson}, {Andrews}, {Aquino-Ort{\'\i}z}, {Arag{\'o}n-Salamanca}, {Argudo-Fern{\'a}ndez}, {Armengaud}, {Aubourg}, {Avila-Reese}, {Badenes}, {Bailey}, {Barger}, {Barrera-Ballesteros}, {Bartosz}, {Bates}, {Baumgarten}, {Bautista}, {Beaton}, {Beers}, {Belfiore}, {Bender}, {Berlind}, {Bernardi}, {Beutler}, {Bird}, {Bizyaev}, {Blanc}, {Blomqvist}, {Bolton}, {Boquien}, {Borissova}, {van den Bosch}, {Bovy}, {Brandt}, {Brinkmann}, {Brownstein}, {Bundy}, {Burgasser}, {Burtin}, {Busca}, {Cappellari}, {Delgado Carigi}, {Carlberg}, {Carnero Rosell}, {Carrera}, {Chanover}, {Cherinka}, {Cheung}, {G{\'o}mez Maqueo Chew}, {Chiappini}, {Choi}, {Chojnowski}, {Chuang}, {Chung}, {Cirolini}, {Clerc}, {Cohen}, {Comparat}, {da Costa}, {Cousinou}, {Covey}, {Crane}, {Croft}, {Cruz-Gonzalez}, {Garrido Cuadra}, {Cunha}, {Damke}, {Darling}, {Davies}, {Dawson}, {de la Macorra}, {Dell'Agli}, {De
  Lee}, {Delubac}, {Di Mille}, {Diamond-Stanic}, {Cano-D{\'\i}az}, {Donor}, {Downes}, {Drory}, {du Mas des Bourboux}, {Duckworth}, {Dwelly}, {Dyer}, {Ebelke}, {Eigenbrot}, {Eisenstein}, {Emsellem}, {Eracleous}, {Escoffier}, {Evans}, {Fan}, {Fern{\'a}ndez-Alvar}, {Fernandez-Trincado}, {Feuillet}, {Finoguenov}, {Fleming}, {Font-Ribera}, {Fredrickson}, {Freischlad}, {Frinchaboy}, {Fuentes}, {Galbany}, {Garcia-Dias}, {Garc{\'\i}a-Hern{\'a}ndez}, {Gaulme}, {Geisler}, {Gelfand}, {Gil-Mar{\'\i}n}, {Gillespie}, {Goddard}, {Gonzalez-Perez}, {Grabowski}, {Green}, {Grier}, {Gunn}, {Guo}, {Guy}, {Hagen}, {Hahn}, {Hall}, {Harding}, {Hasselquist}, {Hawley}, {Hearty}, {Gonzalez Hern{\'a}ndez}, {Ho}, {Hogg}, {Holley-Bockelmann}, {Holtzman}, {Holzer}, {Huehnerhoff}, {Hutchinson}, {Hwang}, {Ibarra-Medel}, {da Silva Ilha}, {Ivans}, {Ivory}, {Jackson}, {Jensen}, {Johnson}, {Jones}, {J{\"o}nsson}, {Jullo}, {Kamble}, {Kinemuchi}, {Kirkby}, {Kitaura}, {Klaene}, {Knapp}, {Kneib}, {Kollmeier}, {Lacerna}, {Lane}, {Lang}, {Law},
  {Lazarz}, {Lee}, {Le Goff}, {Liang}, {Li}, {Li}, {Lian}, {Lima}, {Lin}, {Lin}, {Bertran de Lis}, {Liu}, {de Icaza Lizaola}, {Long}, {Lucatello}, {Lundgren}, {MacDonald}, {Deconto Machado}, {MacLeod}, {Mahadevan}, {Geimba Maia}, {Maiolino}, {Majewski}, {Malanushenko}, {Malanushenko}, {Manchado}, {Mao}, {Maraston}, {Marques-Chaves}, {Masseron}, {Masters}, {McBride}, {McDermid}, {McGrath}, {McGreer}, {Medina Pe{\~n}a}, {Melendez}, {Merloni}, {Merrifield}, {Meszaros}, {Meza}, {Minchev}, {Minniti}, {Miyaji}, {More}, {Mulchaey}, {M{\"u}ller-S{\'a}nchez}, {Muna}, {Munoz}, {Myers}, {Nair}, {Nandra}, {Correa do Nascimento}, {Negrete}, {Ness}, {Newman}, {Nichol}, {Nidever}, {Nitschelm}, {Ntelis}, {O'Connell}, {Oelkers}, {Oravetz}, {Oravetz}, {Pace}, {Padilla}, {Palanque-Delabrouille}, {Alonso Palicio}, {Pan}, {Parejko}, {Parikh}, {P{\^a}ris}, {Park}, {Patten}, {Peirani}, {Pellejero-Ibanez}, {Penny}, {Percival}, {Perez-Fournon}, {Petitjean}, {Pieri}, {Pinsonneault}, {Pisani}, {Poleski}, {Prada}, {Prakash}, {Queiroz},
  {Raddick}, {Raichoor}, {Barboza Rembold}, {Richstein}, {Riffel}, {Riffel}, {Rix}, {Robin}, {Rockosi}, {Rodr{\'\i}guez-Torres}, {Roman-Lopes}, {Rom{\'a}n-Z{\'u}{\~n}iga}, {Rosado}, {Ross}, {Rossi}, {Ruan}, {Ruggeri}, {Rykoff}, {Salazar-Albornoz}, {Salvato}, {S{\'a}nchez}, {Aguado}, {S{\'a}nchez-Gallego}, {Santana}, {Santiago}, {Sayres}, {Schiavon}, {da Silva Schimoia}, {Schlafly}, {Schlegel}, {Schneider}, {Schultheis}, {Schuster}, {Schwope}, {Seo}, {Shao}, {Shen}, {Shetrone}, {Shull}, {Simon}, {Skinner}, {Skrutskie}, {Slosar}, {Smith}, {Sobeck}, {Sobreira}, {Somers}, {Souto}, {Stark}, {Stassun}, {Stauffer}, {Steinmetz}, {Storchi-Bergmann}, {Streblyanska}, {Stringfellow}, {Su{\'a}rez}, {Sun}, {Suzuki}, {Szigeti}, {Taghizadeh-Popp}, {Tang}, {Tao}, {Tayar}, {Tembe}, {Teske}, {Thakar}, {Thomas}, {Thompson}, {Tinker}, {Tissera}, {Tojeiro}, {Hernandez Toledo}, {de la Torre}, {Tremonti}, {Troup}, {Valenzuela}, {Martinez Valpuesta}, {Vargas-Gonz{\'a}lez}, {Vargas-Maga{\~n}a}, {Vazquez}, {Villanova}, {Vivek}, {Vogt},
  {Wake}, {Walterbos}, {Wang}, {Weaver}, {Weijmans}, {Weinberg}, {Westfall}, {Whelan}, {Wild}, {Wilson}, {Wood-Vasey}, {Wylezalek}, {Xiao}, {Yan}, {Yang}, {Ybarra}, {Y{\`e}che}, {Zakamska}, {Zamora}, {Zarrouk}, {Zasowski}, {Zhang}, {Zhao}, {Zheng}, {Zheng}, {Zhou}, {Zhou}, {Zhu}, {Zoccali}, \& {Zou}}]{blanton2017}
{Blanton}, M.~R., {Bershady}, M.~A., {Abolfathi}, B., {et~al.} 2017, \href{http://dx.doi.org/10.3847/1538-3881/aa7567}{\JournalTitle{\aj}, 154, 28}

\bibitem[{{Bowen} \& {Vaughan}(1973)}]{bowen1973}
{Bowen}, I.~S., \& {Vaughan}, A.~H., J. 1973, \href{http://dx.doi.org/10.1364/AO.12.001430}{\JournalTitle{\ao}, 12, 1430}

\bibitem[{{Bressan} {et~al.}(2012){Bressan}, {Marigo}, {Girardi}, {Salasnich}, {Dal Cero}, {Rubele}, \& {Nanni}}]{Bressan2012_parsec}
{Bressan}, A., {Marigo}, P., {Girardi}, L., {et~al.} 2012, \href{http://dx.doi.org/10.1111/j.1365-2966.2012.21948.x}{\JournalTitle{\mnras}, 427, 127}

\bibitem[{{Cantat-Gaudin} {et~al.}(2020){Cantat-Gaudin}, {Anders}, {Castro-Ginard}, {Jordi}, {Romero-G{\'o}mez}, {Soubiran}, {Casamiquela}, {Tarricq}, {Moitinho}, {Vallenari}, {Bragaglia}, {Krone-Martins}, \& {Kounkel}}]{CANTATGAUDIN_2020}
{Cantat-Gaudin}, T., {Anders}, F., {Castro-Ginard}, A., {et~al.} 2020, \href{http://dx.doi.org/10.1051/0004-6361/202038192}{\JournalTitle{\aap}, 640, A1}

\bibitem[{{Cao} \& {Pinsonneault}(2022)}]{cao2022}
{Cao}, L., \& {Pinsonneault}, M.~H. 2022, \href{http://dx.doi.org/10.1093/mnras/stac2706}{\JournalTitle{\mnras}, 517, 2165}

\bibitem[{{Chabrier} {et~al.}(2007){Chabrier}, {Gallardo}, \& {Baraffe}}]{chabrier2007}
{Chabrier}, G., {Gallardo}, J., \& {Baraffe}, I. 2007, \href{http://dx.doi.org/10.1051/0004-6361:20077702}{\JournalTitle{\aap}, 472, L17}

\bibitem[{{Choi} {et~al.}(2016){Choi}, {Dotter}, {Conroy}, {Cantiello}, {Paxton}, \& {Johnson}}]{choi2016_MIST}
{Choi}, J., {Dotter}, A., {Conroy}, C., {et~al.} 2016, \href{http://dx.doi.org/10.3847/0004-637X/823/2/102}{\JournalTitle{\apj}, 823, 102}

\bibitem[{{Covey} {et~al.}(2016){Covey}, {Ag{\"u}eros}, {Law}, {Liu}, {Ahmadi}, {Laher}, {Levitan}, {Sesar}, \& {Surace}}]{covey2016}
{Covey}, K.~R., {Ag{\"u}eros}, M.~A., {Law}, N.~M., {et~al.} 2016, \href{http://dx.doi.org/10.3847/0004-637X/822/2/81}{\JournalTitle{\apj}, 822, 81}

\bibitem[{{Cristofari} {et~al.}(2023{\natexlab{a}}){Cristofari}, {Donati}, {Folsom}, {Masseron}, {Fouqu{\'e}}, {Moutou}, {Artigau}, {Carmona}, {Petit}, {Delfosse}, {Martioli}, \& {the SLS consortium}}]{cristofari2023a}
{Cristofari}, P.~I., {Donati}, J.~F., {Folsom}, C.~P., {et~al.} 2023{\natexlab{a}}, \href{http://dx.doi.org/10.1093/mnras/stad865}{\JournalTitle{\mnras}, 522, 1342}

\bibitem[{{Cristofari} {et~al.}(2023{\natexlab{b}}){Cristofari}, {Donati}, {Moutou}, {Lehmann}, {Charpentier}, {Fouqu{\'e}}, {Folsom}, {Masseron}, {Carmona}, {Delfosse}, {Petit}, {Artigau}, {Cook}, \& {SLS Consortium}}]{cristofari2023b}
{Cristofari}, P.~I., {Donati}, J.~F., {Moutou}, C., {et~al.} 2023{\natexlab{b}}, \href{http://dx.doi.org/10.1093/mnras/stad3144}{\JournalTitle{\mnras}, 526, 5648}

\bibitem[{{Cui} {et~al.}(2012){Cui}, {Zhao}, {Chu}, {Li}, {Li}, {Zhang}, {Su}, {Yao}, {Wang}, {Xing}, {Li}, {Zhu}, {Wang}, {Gu}, {Luo}, {Xu}, {Zhang}, {Liu}, {Zhang}, {Yang}, {Cao}, {Chen}, {Chen}, {Chen}, {Chen}, {Chu}, {Feng}, {Gong}, {Hou}, {Hu}, {Hu}, {Hu}, {Jia}, {Jiang}, {Jiang}, {Jiang}, {Jin}, {Li}, {Li}, {Li}, {Liu}, {Liu}, {Lu}, {Mao}, {Men}, {Qi}, {Qi}, {Shi}, {Tang}, {Tao}, {Wang}, {Wang}, {Wang}, {Wang}, {Wang}, {Wang}, {Wang}, {Wang}, {Wang}, {Wang}, {Wang}, {Wang}, {Xu}, {Xu}, {Yang}, {Yu}, {Yuan}, {Yuan}, {Zhai}, {Zhang}, {Zhang}, {Zhang}, {Zhao}, {Zhou}, {Zhou}, {Zhu}, \& {Zou}}]{cui2012}
{Cui}, X.-Q., {Zhao}, Y.-H., {Chu}, Y.-Q., {et~al.} 2012, \href{http://dx.doi.org/10.1088/1674-4527/12/9/003}{\JournalTitle{Research in Astronomy and Astrophysics}, 12, 1197}

\bibitem[{{Dahm}(2015)}]{dahm2015_age4}
{Dahm}, S.~E. 2015, \href{http://dx.doi.org/10.1088/0004-637X/813/2/108}{\JournalTitle{\apj}, 813, 108}

\bibitem[{{Dotter} {et~al.}(2008){Dotter}, {Chaboyer}, {Jevremovi{\'c}}, {Kostov}, {Baron}, \& {Ferguson}}]{Dotter2008_dartmouth}
{Dotter}, A., {Chaboyer}, B., {Jevremovi{\'c}}, D., {et~al.} 2008, \href{http://dx.doi.org/10.1086/589654}{\JournalTitle{\apjs}, 178, 89}

\bibitem[{{Fang} {et~al.}(2018){Fang}, {Zhao}, {Zhao}, \& {Bharat Kumar}}]{fang2018}
{Fang}, X.-S., {Zhao}, G., {Zhao}, J.-K., \& {Bharat Kumar}, Y. 2018, \href{http://dx.doi.org/10.1093/mnras/sty212}{\JournalTitle{\mnras}, 476, 908}

\bibitem[{{Feiden} {et~al.}(2015){Feiden}, {Jones}, \& {Chaboyer}}]{feiden2015}
{Feiden}, G.~A., {Jones}, J., \& {Chaboyer}, B. 2015, in Cambridge Workshop on Cool Stars, Stellar Systems, and the Sun, Vol.~18, 18th Cambridge Workshop on Cool Stars, Stellar Systems, and the Sun, 171

\bibitem[{{Foreman-Mackey} {et~al.}(2013){Foreman-Mackey}, {Conley}, {Meierjurgen Farr}, {Hogg}, {Lang}, {Marshall}, {Price-Whelan}, {Sanders}, \& {Zuntz}}]{emcee2013}
{Foreman-Mackey}, D., {Conley}, A., {Meierjurgen Farr}, W., {et~al.} 2013, {emcee: The MCMC Hammer}, Astrophysics Source Code Library, record ascl:1303.002

\bibitem[{{Gaia Collaboration}(2022)}]{gaia2022}
{Gaia Collaboration}. 2022, {VizieR Online Data Catalog: Gaia DR3 Part 1. Main source (Gaia Collaboration, 2022)}, VizieR On-line Data Catalog: I/355. Originally published in: Astron. Astrophys., in prep. (2022)

\bibitem[{{Garc{\'\i}a P{\'e}rez} {et~al.}(2016){Garc{\'\i}a P{\'e}rez}, {Allende Prieto}, {Holtzman}, {Shetrone}, {M{\'e}sz{\'a}ros}, {Bizyaev}, {Carrera}, {Cunha}, {Garc{\'\i}a-Hern{\'a}ndez}, {Johnson}, {Majewski}, {Nidever}, {Schiavon}, {Shane}, {Smith}, {Sobeck}, {Troup}, {Zamora}, {Weinberg}, {Bovy}, {Eisenstein}, {Feuillet}, {Frinchaboy}, {Hayden}, {Hearty}, {Nguyen}, {O'Connell}, {Pinsonneault}, {Wilson}, \& {Zasowski}}]{garciaperez2016_aspcap}
{Garc{\'\i}a P{\'e}rez}, A.~E., {Allende Prieto}, C., {Holtzman}, J.~A., {et~al.} 2016, \href{http://dx.doi.org/10.3847/0004-6256/151/6/144}{\JournalTitle{\aj}, 151, 144}

\bibitem[{{Gunn} {et~al.}(2006){Gunn}, {Siegmund}, {Mannery}, {Owen}, {Hull}, {Leger}, {Carey}, {Knapp}, {York}, {Boroski}, {Kent}, {Lupton}, {Rockosi}, {Evans}, {Waddell}, {Anderson}, {Annis}, {Barentine}, {Bartoszek}, {Bastian}, {Bracker}, {Brewington}, {Briegel}, {Brinkmann}, {Brown}, {Carr}, {Czarapata}, {Drennan}, {Dombeck}, {Federwitz}, {Gillespie}, {Gonzales}, {Hansen}, {Harvanek}, {Hayes}, {Jordan}, {Kinney}, {Klaene}, {Kleinman}, {Kron}, {Kresinski}, {Lee}, {Limmongkol}, {Lindenmeyer}, {Long}, {Loomis}, {McGehee}, {Mantsch}, {Neilsen}, {Neswold}, {Newman}, {Nitta}, {Peoples}, {Pier}, {Prieto}, {Prosapio}, {Rivetta}, {Schneider}, {Snedden}, \& {Wang}}]{gunn2006_sdss}
{Gunn}, J.~E., {Siegmund}, W.~A., {Mannery}, E.~J., {et~al.} 2006, \href{http://dx.doi.org/10.1086/500975}{\JournalTitle{\aj}, 131, 2332}

\bibitem[{{Gustafsson} {et~al.}(2008){Gustafsson}, {Edvardsson}, {Eriksson}, {J{\o}rgensen}, {Nordlund}, \& {Plez}}]{gustafsson2008_marcs}
{Gustafsson}, B., {Edvardsson}, B., {Eriksson}, K., {et~al.} 2008, \href{http://dx.doi.org/10.1051/0004-6361:200809724}{\JournalTitle{\aap}, 486, 951}

\bibitem[{{Hahlin} \& {Kochukhov}(2022)}]{hahlin2022}
{Hahlin}, A., \& {Kochukhov}, O. 2022, \href{http://dx.doi.org/10.1051/0004-6361/202142425}{\JournalTitle{\aap}, 659, A151}

\bibitem[{{Hahlin} {et~al.}(2021){Hahlin}, {Kochukhov}, {Alecian}, {Morin}, \& {BinaMIcS Collaboration}}]{hahlin2021}
{Hahlin}, A., {Kochukhov}, O., {Alecian}, E., {Morin}, J., \& {BinaMIcS Collaboration}. 2021, \href{http://dx.doi.org/10.1051/0004-6361/202140832}{\JournalTitle{\aap}, 650, A197}

\bibitem[{{Hahlin} {et~al.}(2023){Hahlin}, {Kochukhov}, {Rains}, {Lavail}, {Hatzes}, {Piskunov}, {Reiners}, {Seemann}, {Boldt-Christmas}, {Guenther}, {Heiter}, {Nortmann}, {Yan}, {Shulyak}, {Smoker}, {Rodler}, {Bristow}, {Dorn}, {Jung}, {Marquart}, \& {Stempels}}]{hahlin2023}
{Hahlin}, A., {Kochukhov}, O., {Rains}, A.~D., {et~al.} 2023, \href{http://dx.doi.org/10.1051/0004-6361/202346314}{\JournalTitle{\aap}, 675, A91}

\bibitem[{{Han} {et~al.}(2023){Han}, {L{\'o}pez-Valdivia}, {Mace}, \& {Jaffe}}]{eunkyu2023}
{Han}, E., {L{\'o}pez-Valdivia}, R., {Mace}, G.~N., \& {Jaffe}, D.~T. 2023, \href{http://dx.doi.org/10.3847/1538-3881/acd2dd}{\JournalTitle{\aj}, 166, 4}

\bibitem[{Harris {et~al.}(2020)Harris, Millman, van~der Walt, Gommers, Virtanen, Cournapeau, Wieser, Taylor, Berg, Smith, Kern, Picus, Hoyer, van Kerkwijk, Brett, Haldane, del R{\'{i}}o, Wiebe, Peterson, G{\'{e}}rard-Marchant, Sheppard, Reddy, Weckesser, Abbasi, Gohlke, \& Oliphant}]{harris2020_numpy}
Harris, C.~R., Millman, K.~J., van~der Walt, S.~J., {et~al.} 2020, \href{http://dx.doi.org/10.1038/s41586-020-2649-2}{\JournalTitle{Nature}, 585, 357}

\bibitem[{{Hartman} {et~al.}(2010){Hartman}, {Bakos}, {Kov{\'a}cs}, \& {Noyes}}]{hartman2010}
{Hartman}, J.~D., {Bakos}, G.~{\'A}., {Kov{\'a}cs}, G., \& {Noyes}, R.~W. 2010, \href{http://dx.doi.org/10.1111/j.1365-2966.2010.17147.x}{\JournalTitle{\mnras}, 408, 475}

\bibitem[{{Hawley} {et~al.}(2014){Hawley}, {Davenport}, {Kowalski}, {Wisniewski}, {Hebb}, {Deitrick}, \& {Hilton}}]{hawley2014}
{Hawley}, S.~L., {Davenport}, J. R.~A., {Kowalski}, A.~F., {et~al.} 2014, \href{http://dx.doi.org/10.1088/0004-637X/797/2/121}{\JournalTitle{\apj}, 797, 121}

\bibitem[{{Heyl} {et~al.}(2022){Heyl}, {Caiazzo}, \& {Richer}}]{HEYL_2022}
{Heyl}, J., {Caiazzo}, I., \& {Richer}, H.~B. 2022, \href{http://dx.doi.org/10.3847/1538-4357/ac45fc}{\JournalTitle{\apj}, 926, 132}

\bibitem[{{Howell} {et~al.}(2014){Howell}, {Sobeck}, {Haas}, {Still}, {Barclay}, {Mullally}, {Troeltzsch}, {Aigrain}, {Bryson}, {Caldwell}, {Chaplin}, {Cochran}, {Huber}, {Marcy}, {Miglio}, {Najita}, {Smith}, {Twicken}, \& {Fortney}}]{howell2014}
{Howell}, S.~B., {Sobeck}, C., {Haas}, M., {et~al.} 2014, \href{http://dx.doi.org/10.1086/676406}{\JournalTitle{\pasp}, 126, 398}

\bibitem[{{Hubeny} \& {Lanz}(2011)}]{hubeny2011_synspec}
{Hubeny}, I., \& {Lanz}, T. 2011, {Synspec: General Spectrum Synthesis Program}, Astrophysics Source Code Library, record ascl:1109.022

\bibitem[{Hunter(2007)}]{Hunter2007_matplotlib}
Hunter, J.~D. 2007, \href{http://dx.doi.org/10.1109/MCSE.2007.55}{\JournalTitle{Computing in Science \& Engineering}, 9, 90}

\bibitem[{{Jackson} {et~al.}(2018){Jackson}, {Deliyannis}, \& {Jeffries}}]{Jackson2018}
{Jackson}, R.~J., {Deliyannis}, C.~P., \& {Jeffries}, R.~D. 2018, \href{http://dx.doi.org/10.1093/mnras/sty374}{\JournalTitle{\mnras}, 476, 3245}

\bibitem[{{Jackson} {et~al.}(2019){Jackson}, {Jeffries}, {Deliyannis}, {Sun}, \& {Douglas}}]{Jackson2019}
{Jackson}, R.~J., {Jeffries}, R.~D., {Deliyannis}, C.~P., {Sun}, Q., \& {Douglas}, S.~T. 2019, \href{http://dx.doi.org/10.1093/mnras/sty3184}{\JournalTitle{\mnras}, 483, 1125}

\bibitem[{{Jackson} {et~al.}(2016){Jackson}, {Jeffries}, {Randich}, {Bragaglia}, {Carraro}, {Costado}, {Flaccomio}, {Lanzafame}, {Lardo}, {Monaco}, {Morbidelli}, {Smiljanic}, \& {Zaggia}}]{Jackson2016}
{Jackson}, R.~J., {Jeffries}, R.~D., {Randich}, S., {et~al.} 2016, \href{http://dx.doi.org/10.1051/0004-6361/201527507}{\JournalTitle{\aap}, 586, A52}

\bibitem[{{Jaehnig} {et~al.}(2019){Jaehnig}, {Somers}, \& {Stassun}}]{jaehnig2019}
{Jaehnig}, K., {Somers}, G., \& {Stassun}, K.~G. 2019, \href{http://dx.doi.org/10.3847/1538-4357/ab21cf}{\JournalTitle{\apj}, 879, 39}

\bibitem[{{Jeffers} {et~al.}(2018){Jeffers}, {Sch{\"o}fer}, {Lamert}, {Reiners}, {Montes}, {Caballero}, {Cort{\'e}s-Contreras}, {Marvin}, {Passegger}, {Zechmeister}, {Quirrenbach}, {Alonso-Floriano}, {Amado}, {Bauer}, {Casal}, {Diez Alonso}, {Herrero}, {Morales}, {Mundt}, {Ribas}, \& {Sarmiento}}]{jeffers2018}
{Jeffers}, S.~V., {Sch{\"o}fer}, P., {Lamert}, A., {et~al.} 2018, \href{http://dx.doi.org/10.1051/0004-6361/201629599}{\JournalTitle{\aap}, 614, A76}

\bibitem[{{Johns-Krull}(2007)}]{johns-krull2007}
{Johns-Krull}, C.~M. 2007, \href{http://dx.doi.org/10.1086/519017}{\JournalTitle{\apj}, 664, 975}

\bibitem[{{Johns-Krull} \& {Valenti}(1996)}]{johns-krull1996_olegreview15}
{Johns-Krull}, C.~M., \& {Valenti}, J.~A. 1996, \href{http://dx.doi.org/10.1086/309954}{\JournalTitle{\apjl}, 459, L95}

\bibitem[{{Johns-Krull} \& {Valenti}(2000)}]{johns-krull2000_olegreview8}
{Johns-Krull}, C.~M., \& {Valenti}, J.~A. 2000, in Astronomical Society of the Pacific Conference Series, Vol. 198, Stellar Clusters and Associations: Convection, Rotation, and Dynamos, ed. R.~{Pallavicini}, G.~{Micela}, \& S.~{Sciortino}, 371

\bibitem[{{Johns-Krull} {et~al.}(2004){Johns-Krull}, {Valenti}, \& {Saar}}]{johns-krull2004}
{Johns-Krull}, C.~M., {Valenti}, J.~A., \& {Saar}, S.~H. 2004, \href{http://dx.doi.org/10.1086/425652}{\JournalTitle{\apj}, 617, 1204}

\bibitem[{{Kawaler}(1988)}]{kawaler1988}
{Kawaler}, S.~D. 1988, \href{http://dx.doi.org/10.1086/166740}{\JournalTitle{\apj}, 333, 236}

\bibitem[{{Kesseli} {et~al.}(2018){Kesseli}, {Muirhead}, {Mann}, \& {Mace}}]{Kesseli2018}
{Kesseli}, A.~Y., {Muirhead}, P.~S., {Mann}, A.~W., \& {Mace}, G. 2018, \href{http://dx.doi.org/10.3847/1538-3881/aabccb}{\JournalTitle{\aj}, 155, 225}

\bibitem[{{Kochukhov}(2016)}]{kochukhov2016}
{Kochukhov}, O. 2016, \href{http://dx.doi.org/10.1007/978-3-319-24151-7_9}{in Lecture Notes in Physics, Berlin Springer Verlag, ed. J.-P. {Rozelot} \& C.~{Neiner}, Vol. 914}, 177

\bibitem[{{Kochukhov}(2021)}]{kochukhov2021}
---. 2021, \href{http://dx.doi.org/10.1007/s00159-020-00130-3}{\JournalTitle{\aapr}, 29, 1}

\bibitem[{{Kochukhov} {et~al.}(2009){Kochukhov}, {Heiter}, {Piskunov}, {Ryde}, {Gustafsson}, {Bagnulo}, \& {Plez}}]{kochukhov2009_olegreview6}
{Kochukhov}, O., {Heiter}, U., {Piskunov}, N., {et~al.} 2009, \href{http://dx.doi.org/10.1063/1.3099081}{in American Institute of Physics Conference Series, Vol. 1094, 15th Cambridge Workshop on Cool Stars, Stellar Systems, and the Sun, ed. E.~{Stempels}}, 124

\bibitem[{{Kochukhov} \& {Lavail}(2017)}]{kochukhov2017_olegreview4}
{Kochukhov}, O., \& {Lavail}, A. 2017, \href{http://dx.doi.org/10.3847/2041-8213/835/1/L4}{\JournalTitle{\apjl}, 835, L4}

\bibitem[{{Kochukhov} {et~al.}(2010){Kochukhov}, {Makaganiuk}, \& {Piskunov}}]{kochukhov2010_synmast}
{Kochukhov}, O., {Makaganiuk}, V., \& {Piskunov}, N. 2010, \href{http://dx.doi.org/10.1051/0004-6361/201015429}{\JournalTitle{\aap}, 524, A5}

\bibitem[{{Kochukhov} {et~al.}(2001){Kochukhov}, {Piskunov}, {Valenti}, \& {Johns-Krull}}]{kochukhov2001_olegreview5}
{Kochukhov}, O., {Piskunov}, N., {Valenti}, J., \& {Johns-Krull}, C. 2001, in Astronomical Society of the Pacific Conference Series, Vol. 248, Magnetic Fields Across the Hertzsprung-Russell Diagram, ed. G.~{Mathys}, S.~K. {Solanki}, \& D.~T. {Wickramasinghe}, 219

\bibitem[{{Kochukhov} \& {Reiners}(2020)}]{kochukhov2020}
{Kochukhov}, O., \& {Reiners}, A. 2020, \href{http://dx.doi.org/10.3847/1538-4357/abb2a2}{\JournalTitle{\apj}, 902, 43}

\bibitem[{{Kochukhov} \& {Shulyak}(2019)}]{kochukhov2019_olegreview7}
{Kochukhov}, O., \& {Shulyak}, D. 2019, \href{http://dx.doi.org/10.3847/1538-4357/ab06c5}{\JournalTitle{\apj}, 873, 69}

\bibitem[{{Kupka} {et~al.}(1999){Kupka}, {Piskunov}, {Ryabchikova}, {Stempels}, \& {Weiss}}]{kupka1999}
{Kupka}, F., {Piskunov}, N., {Ryabchikova}, T.~A., {Stempels}, H.~C., \& {Weiss}, W.~W. 1999, \href{http://dx.doi.org/10.1051/aas:1999267}{\JournalTitle{\aaps}, 138, 119}

\bibitem[{{Lasker} {et~al.}(2008){Lasker}, {Lattanzi}, {McLean}, {Bucciarelli}, {Drimmel}, {Garcia}, {Greene}, {Guglielmetti}, {Hanley}, {Hawkins}, {Laidler}, {Loomis}, {Meakes}, {Mignani}, {Morbidelli}, {Morrison}, {Pannunzio}, {Rosenberg}, {Sarasso}, {Smart}, {Spagna}, {Sturch}, {Volpicelli}, {White}, {Wolfe}, \& {Zacchei}}]{lasker2008_gsc}
{Lasker}, B.~M., {Lattanzi}, M.~G., {McLean}, B.~J., {et~al.} 2008, \href{http://dx.doi.org/10.1088/0004-6256/136/2/735}{\JournalTitle{\aj}, 136, 735}

\bibitem[{{Lavail} {et~al.}(2019){Lavail}, {Kochukhov}, \& {Hussain}}]{lavail2019}
{Lavail}, A., {Kochukhov}, O., \& {Hussain}, G.~A.~J. 2019, \href{http://dx.doi.org/10.1051/0004-6361/201935695}{\JournalTitle{\aap}, 630, A99}

\bibitem[{{Law} {et~al.}(2009){Law}, {Kulkarni}, {Dekany}, {Ofek}, {Quimby}, {Nugent}, {Surace}, {Grillmair}, {Bloom}, {Kasliwal}, {Bildsten}, {Brown}, {Cenko}, {Ciardi}, {Croner}, {Djorgovski}, {van Eyken}, {Filippenko}, {Fox}, {Gal-Yam}, {Hale}, {Hamam}, {Helou}, {Henning}, {Howell}, {Jacobsen}, {Laher}, {Mattingly}, {McKenna}, {Pickles}, {Poznanski}, {Rahmer}, {Rau}, {Rosing}, {Shara}, {Smith}, {Starr}, {Sullivan}, {Velur}, {Walters}, \& {Zolkower}}]{law2009}
{Law}, N.~M., {Kulkarni}, S.~R., {Dekany}, R.~G., {et~al.} 2009, \href{http://dx.doi.org/10.1086/648598}{\JournalTitle{\pasp}, 121, 1395}

\bibitem[{{Lodieu} {et~al.}(2019){Lodieu}, {P{\'e}rez-Garrido}, {Smart}, \& {Silvotti}}]{lodieu2019}
{Lodieu}, N., {P{\'e}rez-Garrido}, A., {Smart}, R.~L., \& {Silvotti}, R. 2019, \href{http://dx.doi.org/10.1051/0004-6361/201935533}{\JournalTitle{\aap}, 628, A66}

\bibitem[{{Majewski} {et~al.}(2017){Majewski}, {Schiavon}, {Frinchaboy}, {Allende Prieto}, {Barkhouser}, {Bizyaev}, {Blank}, {Brunner}, {Burton}, {Carrera}, {Chojnowski}, {Cunha}, {Epstein}, {Fitzgerald}, {Garc{\'\i}a P{\'e}rez}, {Hearty}, {Henderson}, {Holtzman}, {Johnson}, {Lam}, {Lawler}, {Maseman}, {M{\'e}sz{\'a}ros}, {Nelson}, {Nguyen}, {Nidever}, {Pinsonneault}, {Shetrone}, {Smee}, {Smith}, {Stolberg}, {Skrutskie}, {Walker}, {Wilson}, {Zasowski}, {Anders}, {Basu}, {Beland}, {Blanton}, {Bovy}, {Brownstein}, {Carlberg}, {Chaplin}, {Chiappini}, {Eisenstein}, {Elsworth}, {Feuillet}, {Fleming}, {Galbraith-Frew}, {Garc{\'\i}a}, {Garc{\'\i}a-Hern{\'a}ndez}, {Gillespie}, {Girardi}, {Gunn}, {Hasselquist}, {Hayden}, {Hekker}, {Ivans}, {Kinemuchi}, {Klaene}, {Mahadevan}, {Mathur}, {Mosser}, {Muna}, {Munn}, {Nichol}, {O'Connell}, {Parejko}, {Robin}, {Rocha-Pinto}, {Schultheis}, {Serenelli}, {Shane}, {Silva Aguirre}, {Sobeck}, {Thompson}, {Troup}, {Weinberg}, \& {Zamora}}]{majewski2017_apogee}
{Majewski}, S.~R., {Schiavon}, R.~P., {Frinchaboy}, P.~M., {et~al.} 2017, \href{http://dx.doi.org/10.3847/1538-3881/aa784d}{\JournalTitle{\aj}, 154, 94}

\bibitem[{{Mann} {et~al.}(2015){Mann}, {Feiden}, {Gaidos}, {Boyajian}, \& {von Braun}}]{mann2015}
{Mann}, A.~W., {Feiden}, G.~A., {Gaidos}, E., {Boyajian}, T., \& {von Braun}, K. 2015, \href{http://dx.doi.org/10.1088/0004-637X/804/1/64}{\JournalTitle{\apj}, 804, 64}

\bibitem[{{Mann} {et~al.}(2016){Mann}, {Feiden}, {Gaidos}, {Boyajian}, \& {von Braun}}]{mann2016}
---. 2016, \href{http://dx.doi.org/10.3847/0004-637X/819/1/87}{\JournalTitle{\apj}, 819, 87}

\bibitem[{{Micela} {et~al.}(1996){Micela}, {Sciortino}, {Kashyap}, {Harnden}, \& {Rosner}}]{micela1996}
{Micela}, G., {Sciortino}, S., {Kashyap}, V., {Harnden}, F.~R., J., \& {Rosner}, R. 1996, \href{http://dx.doi.org/10.1086/192252}{\JournalTitle{\apjs}, 102, 75}

\bibitem[{{Micela} {et~al.}(1990){Micela}, {Sciortino}, {Vaiana}, {Harnden}, {Rosner}, \& {Schmitt}}]{micela1990}
{Micela}, G., {Sciortino}, S., {Vaiana}, G.~S., {et~al.} 1990, \href{http://dx.doi.org/10.1086/168263}{\JournalTitle{\apj}, 348, 557}

\bibitem[{{Micela} {et~al.}(1999){Micela}, {Sciortino}, {Harnden}, {Kashyap}, {Rosner}, {Prosser}, {Damiani}, {Stauffer}, \& {Caillault}}]{micela1999}
{Micela}, G., {Sciortino}, S., {Harnden}, F.~R., J., {et~al.} 1999, \JournalTitle{\aap}, 341, 751

\bibitem[{{Muirhead} {et~al.}(2018){Muirhead}, {Dressing}, {Mann}, {Rojas-Ayala}, {L{\'e}pine}, {Paegert}, {De Lee}, \& {Oelkers}}]{muirhead2018_vmag}
{Muirhead}, P.~S., {Dressing}, C.~D., {Mann}, A.~W., {et~al.} 2018, \href{http://dx.doi.org/10.3847/1538-3881/aab710}{\JournalTitle{\aj}, 155, 180}

\bibitem[{{Newton} {et~al.}(2017){Newton}, {Irwin}, {Charbonneau}, {Berlind}, {Calkins}, \& {Mink}}]{newton2017}
{Newton}, E.~R., {Irwin}, J., {Charbonneau}, D., {et~al.} 2017, \href{http://dx.doi.org/10.3847/1538-4357/834/1/85}{\JournalTitle{\apj}, 834, 85}

\bibitem[{{Newton} {et~al.}(2016){Newton}, {Irwin}, {Charbonneau}, {Berta-Thompson}, \& {Dittmann}}]{newton2016}
{Newton}, E.~R., {Irwin}, J., {Charbonneau}, D., {Berta-Thompson}, Z.~K., \& {Dittmann}, J.~A. 2016, \href{http://dx.doi.org/10.3847/2041-8205/821/1/L19}{\JournalTitle{\apjl}, 821, L19}

\bibitem[{{Nguyen} {et~al.}(2022){Nguyen}, {Costa}, {Girardi}, {Volpato}, {Bressan}, {Chen}, {Marigo}, {Fu}, \& {Goudfrooij}}]{nguyen2022}
{Nguyen}, C.~T., {Costa}, G., {Girardi}, L., {et~al.} 2022, \href{http://dx.doi.org/10.1051/0004-6361/202244166}{\JournalTitle{\aap}, 665, A126}

\bibitem[{{Nidever} {et~al.}(2015){Nidever}, {Holtzman}, {Allende Prieto}, {Beland}, {Bender}, {Bizyaev}, {Burton}, {Desphande}, {Fleming}, {Garc{\'\i}a P{\'e}rez}, {Hearty}, {Majewski}, {M{\'e}sz{\'a}ros}, {Muna}, {Nguyen}, {Schiavon}, {Shetrone}, {Skrutskie}, {Sobeck}, \& {Wilson}}]{nidever2015}
{Nidever}, D.~L., {Holtzman}, J.~A., {Allende Prieto}, C., {et~al.} 2015, \href{http://dx.doi.org/10.1088/0004-6256/150/6/173}{\JournalTitle{\aj}, 150, 173}

\bibitem[{{N{\'u}{\~n}ez} \& {Ag{\"u}eros}(2016)}]{nunesagueros2016}
{N{\'u}{\~n}ez}, A., \& {Ag{\"u}eros}, M.~A. 2016, \href{http://dx.doi.org/10.3847/0004-637X/830/1/44}{\JournalTitle{\apj}, 830, 44}

\bibitem[{{Perryman} {et~al.}(1998){Perryman}, {Brown}, {Lebreton}, {Gomez}, {Turon}, {Cayrel de Strobel}, {Mermilliod}, {Robichon}, {Kovalevsky}, \& {Crifo}}]{perryman1998}
{Perryman}, M.~A.~C., {Brown}, A.~G.~A., {Lebreton}, Y., {et~al.} 1998, \JournalTitle{\aap}, 331, 81

\bibitem[{{Phan-Bao} {et~al.}(2009){Phan-Bao}, {Lim}, {Donati}, {Johns-Krull}, \& {Mart{\'\i}n}}]{phan-bao_olegreview13}
{Phan-Bao}, N., {Lim}, J., {Donati}, J.-F., {Johns-Krull}, C.~M., \& {Mart{\'\i}n}, E.~L. 2009, \href{http://dx.doi.org/10.1088/0004-637X/704/2/1721}{\JournalTitle{\apj}, 704, 1721}

\bibitem[{{Piskunov} {et~al.}(1995){Piskunov}, {Kupka}, {Ryabchikova}, {Weiss}, \& {Jeffery}}]{piskunov1995}
{Piskunov}, N.~E., {Kupka}, F., {Ryabchikova}, T.~A., {Weiss}, W.~W., \& {Jeffery}, C.~S. 1995, \JournalTitle{\aaps}, 112, 525

\bibitem[{{Plez}(2012)}]{plez2012_turbospectrum}
{Plez}, B. 2012, {Turbospectrum: Code for spectral synthesis}

\bibitem[{{Pouilly} {et~al.}(2023){Pouilly}, {Kochukhov}, {K{\'o}sp{\'a}l}, {Hahlin}, {Carmona}, \& {{\'A}brah{\'a}m}}]{pouilly2023}
{Pouilly}, K., {Kochukhov}, O., {K{\'o}sp{\'a}l}, {\'A}., {et~al.} 2023, \href{http://dx.doi.org/10.1093/mnras/stac3322}{\JournalTitle{\mnras}, 518, 5072}

\bibitem[{{Rau} {et~al.}(2009){Rau}, {Kulkarni}, {Law}, {Bloom}, {Ciardi}, {Djorgovski}, {Fox}, {Gal-Yam}, {Grillmair}, {Kasliwal}, {Nugent}, {Ofek}, {Quimby}, {Reach}, {Shara}, {Bildsten}, {Cenko}, {Drake}, {Filippenko}, {Helfand}, {Helou}, {Howell}, {Poznanski}, \& {Sullivan}}]{rau2009}
{Rau}, A., {Kulkarni}, S.~R., {Law}, N.~M., {et~al.} 2009, \href{http://dx.doi.org/10.1086/605911}{\JournalTitle{\pasp}, 121, 1334}

\bibitem[{{Rebull} {et~al.}(2016){Rebull}, {Stauffer}, {Bouvier}, {Cody}, {Hillenbrand}, {Soderblom}, {Valenti}, {Barrado}, {Bouy}, {Ciardi}, {Pinsonneault}, {Stassun}, {Micela}, {Aigrain}, {Vrba}, {Somers}, {Gillen}, \& {Collier Cameron}}]{rebull2016}
{Rebull}, L.~M., {Stauffer}, J.~R., {Bouvier}, J., {et~al.} 2016, \href{http://dx.doi.org/10.3847/0004-6256/152/5/114}{\JournalTitle{\aj}, 152, 114}

\bibitem[{{Reid} \& {Gizis}(1997)}]{reid1997}
{Reid}, I.~N., \& {Gizis}, J.~E. 1997, \href{http://dx.doi.org/10.1086/118436}{\JournalTitle{\aj}, 113, 2246}

\bibitem[{{Reiners} {et~al.}(2009){Reiners}, {Basri}, \& {Christensen}}]{reiners2009b}
{Reiners}, A., {Basri}, G., \& {Christensen}, U.~R. 2009, \href{http://dx.doi.org/10.1088/0004-637X/697/1/373}{\JournalTitle{\apj}, 697, 373}

\bibitem[{{Reiners} {et~al.}(2012){Reiners}, {Joshi}, \& {Goldman}}]{reiners2012}
{Reiners}, A., {Joshi}, N., \& {Goldman}, B. 2012, \href{http://dx.doi.org/10.1088/0004-6256/143/4/93}{\JournalTitle{\aj}, 143, 93}

\bibitem[{{Reiners} {et~al.}(2022){Reiners}, {Shulyak}, {K{\"a}pyl{\"a}}, {Ribas}, {Nagel}, {Zechmeister}, {Caballero}, {Shan}, {Fuhrmeister}, {Quirrenbach}, {Amado}, {Montes}, {Jeffers}, {Azzaro}, {B{\'e}jar}, {Chaturvedi}, {Henning}, {K{\"u}rster}, \& {Pall{\'e}}}]{reiners2022}
{Reiners}, A., {Shulyak}, D., {K{\"a}pyl{\"a}}, P.~J., {et~al.} 2022, \href{http://dx.doi.org/10.1051/0004-6361/202243251}{\JournalTitle{\aap}, 662, A41}

\bibitem[{{Saar}(1994)}]{saar1994_olegreview12}
{Saar}, S.~H. 1994, in Infrared Solar Physics, ed. D.~M. {Rabin}, J.~T. {Jefferies}, \& C.~{Lindsey}, Vol. 154, 493

\bibitem[{{Saar}(1996)}]{saar1996_olegreview14}
{Saar}, S.~H. 1996, in Stellar Surface Structure, ed. K.~G. {Strassmeier} \& J.~L. {Linsky}, Vol. 176, 237

\bibitem[{{Saar} \& {Linsky}(1985)}]{saar1985_olegreview11}
{Saar}, S.~H., \& {Linsky}, J.~L. 1985, \href{http://dx.doi.org/10.1086/184578}{\JournalTitle{\apjl}, 299, L47}

\bibitem[{{Salpeter}(1955)}]{salpeter1955}
{Salpeter}, E.~E. 1955, \href{http://dx.doi.org/10.1086/145971}{\JournalTitle{\apj}, 121, 161}

\bibitem[{{Shulyak} {et~al.}(2017){Shulyak}, {Reiners}, {Engeln}, {Malo}, {Yadav}, {Morin}, \& {Kochukhov}}]{shulyak2017_olegreview2}
{Shulyak}, D., {Reiners}, A., {Engeln}, A., {et~al.} 2017, \href{http://dx.doi.org/10.1038/s41550-017-0184}{\JournalTitle{Nature Astronomy}, 1, 0184}

\bibitem[{{Shulyak} {et~al.}(2014){Shulyak}, {Reiners}, {Seemann}, {Kochukhov}, \& {Piskunov}}]{shulyak2014_olegreview10}
{Shulyak}, D., {Reiners}, A., {Seemann}, U., {Kochukhov}, O., \& {Piskunov}, N. 2014, \href{http://dx.doi.org/10.1051/0004-6361/201322136}{\JournalTitle{\aap}, 563, A35}

\bibitem[{{Shulyak} {et~al.}(2011){Shulyak}, {Seifahrt}, {Reiners}, {Kochukhov}, \& {Piskunov}}]{shulyak2011_olegreview1}
{Shulyak}, D., {Seifahrt}, A., {Reiners}, A., {Kochukhov}, O., \& {Piskunov}, N. 2011, \href{http://dx.doi.org/10.1111/j.1365-2966.2011.19644.x}{\JournalTitle{\mnras}, 418, 2548}

\bibitem[{{Shulyak} {et~al.}(2019){Shulyak}, {Reiners}, {Nagel}, {Tal-Or}, {Caballero}, {Zechmeister}, {B{\'e}jar}, {Cort{\'e}s-Contreras}, {Martin}, {Kaminski}, {Ribas}, {Quirrenbach}, {Amado}, {Anglada-Escud{\'e}}, {Bauer}, {Dreizler}, {Guenther}, {Henning}, {Jeffers}, {K{\"u}rster}, {Lafarga}, {Montes}, {Morales}, \& {Pedraz}}]{shulyak2019_olegreview3}
{Shulyak}, D., {Reiners}, A., {Nagel}, E., {et~al.} 2019, \href{http://dx.doi.org/10.1051/0004-6361/201935315}{\JournalTitle{\aap}, 626, A86}

\bibitem[{{Skrutskie} {et~al.}(2006){Skrutskie}, {Cutri}, {Stiening}, {Weinberg}, {Schneider}, {Carpenter}, {Beichman}, {Capps}, {Chester}, {Elias}, {Huchra}, {Liebert}, {Lonsdale}, {Monet}, {Price}, {Seitzer}, {Jarrett}, {Kirkpatrick}, {Gizis}, {Howard}, {Evans}, {Fowler}, {Fullmer}, {Hurt}, {Light}, {Kopan}, {Marsh}, {McCallon}, {Tam}, {Van Dyk}, \& {Wheelock}}]{skrustkie2006}
{Skrutskie}, M.~F., {Cutri}, R.~M., {Stiening}, R., {et~al.} 2006, \href{http://dx.doi.org/10.1086/498708}{\JournalTitle{\aj}, 131, 1163}

\bibitem[{{Skumanich}(1972)}]{skumanich1972}
{Skumanich}, A. 1972, \href{http://dx.doi.org/10.1086/151310}{\JournalTitle{\apj}, 171, 565}

\bibitem[{{Smith} {et~al.}(2021){Smith}, {Bizyaev}, {Cunha}, {Shetrone}, {Souto}, {Allende Prieto}, {Masseron}, {M{\'e}sz{\'a}ros}, {J{\"o}nsson}, {Hasselquist}, {Osorio}, {Garc{\'\i}a-Hern{\'a}ndez}, {Plez}, {Beaton}, {Holtzman}, {Majewski}, {Stringfellow}, \& {Sobeck}}]{smith2021}
{Smith}, V.~V., {Bizyaev}, D., {Cunha}, K., {et~al.} 2021, \href{http://dx.doi.org/10.3847/1538-3881/abefdc}{\JournalTitle{\aj}, 161, 254}

\bibitem[{{Soderblom} {et~al.}(2009){Soderblom}, {Laskar}, {Valenti}, {Stauffer}, \& {Rebull}}]{soderblom2009}
{Soderblom}, D.~R., {Laskar}, T., {Valenti}, J.~A., {Stauffer}, J.~R., \& {Rebull}, L.~M. 2009, \href{http://dx.doi.org/10.1088/0004-6256/138/5/1292}{\JournalTitle{\aj}, 138, 1292}

\bibitem[{{Somers} {et~al.}(2020){Somers}, {Cao}, \& {Pinsonneault}}]{Somers2020_spots}
{Somers}, G., {Cao}, L., \& {Pinsonneault}, M.~H. 2020, \href{http://dx.doi.org/10.3847/1538-4357/ab722e}{\JournalTitle{\apj}, 891, 29}

\bibitem[{{Souto} {et~al.}(2021){Souto}, {Cunha}, \& {Smith}}]{souto2021}
{Souto}, D., {Cunha}, K., \& {Smith}, V.~V. 2021, \href{http://dx.doi.org/10.3847/1538-4357/abfdb5}{\JournalTitle{\apj}, 917, 11}

\bibitem[{{Souto} {et~al.}(2017){Souto}, {Cunha}, {Garc{\'\i}a-Hern{\'a}ndez}, {Zamora}, {Allende Prieto}, {Smith}, {Mahadevan}, {Blake}, {Johnson}, {J{\"o}nsson}, {Pinsonneault}, {Holtzman}, {Majewski}, {Shetrone}, {Teske}, {Nidever}, {Schiavon}, {Sobeck}, {Garc{\'\i}a P{\'e}rez}, {G{\'o}mez Maqueo Chew}, \& {Stassun}}]{souto2017}
{Souto}, D., {Cunha}, K., {Garc{\'\i}a-Hern{\'a}ndez}, D.~A., {et~al.} 2017, \href{http://dx.doi.org/10.3847/1538-4357/835/2/239}{\JournalTitle{\apj}, 835, 239}

\bibitem[{{Souto} {et~al.}(2020){Souto}, {Cunha}, {Smith}, {Allende Prieto}, {Burgasser}, {Covey}, {Garc{\'\i}a-Hern{\'a}ndez}, {Holtzman}, {Johnson}, {J{\"o}nsson}, {Mahadevan}, {Majewski}, {Masseron}, {Shetrone}, {Rojas-Ayala}, {Sobeck}, {Stassun}, {Terrien}, {Teske}, {Wanderley}, \& {Zamora}}]{souto2020}
{Souto}, D., {Cunha}, K., {Smith}, V.~V., {et~al.} 2020, \href{http://dx.doi.org/10.3847/1538-4357/ab6d07}{\JournalTitle{\apj}, 890, 133}

\bibitem[{{Stassun} {et~al.}(2012){Stassun}, {Kratter}, {Scholz}, \& {Dupuy}}]{stassun2012}
{Stassun}, K.~G., {Kratter}, K.~M., {Scholz}, A., \& {Dupuy}, T.~J. 2012, \href{http://dx.doi.org/10.1088/0004-637X/756/1/47}{\JournalTitle{\apj}, 756, 47}

\bibitem[{{Stauffer} {et~al.}(1994){Stauffer}, {Caillault}, {Gagne}, {Prosser}, \& {Hartmann}}]{stauffer1994}
{Stauffer}, J.~R., {Caillault}, J.~P., {Gagne}, M., {Prosser}, C.~F., \& {Hartmann}, L.~W. 1994, \href{http://dx.doi.org/10.1086/191951}{\JournalTitle{\apjs}, 91, 625}

\bibitem[{{Stauffer} {et~al.}(2007){Stauffer}, {Hartmann}, {Fazio}, {Allen}, {Patten}, {Lowrance}, {Hurt}, {Rebull}, {Cutri}, {Ramirez}, {Young}, {Rieke}, {Gorlova}, {Muzerolle}, {Slesnick}, \& {Skrutskie}}]{stauffer2007_vmag}
{Stauffer}, J.~R., {Hartmann}, L.~W., {Fazio}, G.~G., {et~al.} 2007, \href{http://dx.doi.org/10.1086/518961}{\JournalTitle{\apjs}, 172, 663}

\bibitem[{{Stelzer} {et~al.}(2000){Stelzer}, {Neuh{\"a}user}, \& {Hambaryan}}]{stelzer2000}
{Stelzer}, B., {Neuh{\"a}user}, R., \& {Hambaryan}, V. 2000, \href{http://dx.doi.org/10.48550/arXiv.astro-ph/0002354}{\JournalTitle{\aap}, 356, 949}

\bibitem[{{Stift} \& {Leone}(2003)}]{stift2003}
{Stift}, M.~J., \& {Leone}, F. 2003, \href{http://dx.doi.org/10.1051/0004-6361:20021605}{\JournalTitle{\aap}, 398, 411}

\bibitem[{{Wanderley} {et~al.}(2023){Wanderley}, {Cunha}, {Souto}, {Smith}, {Cao}, {Pinsonneault}, {Allende Prieto}, {Covey}, {Masseron}, {Pascucci}, {Stassun}, {Terrien}, {Bergsten}, {Bizyaev}, {Fern{\'a}ndez-Trincado}, {J{\"o}nsson}, {Hasselquist}, {Holtzman}, {Lane}, {Mahadevan}, {Majewski}, {Minniti}, {Pan}, {Serna}, {Sobeck}, \& {Stringfellow}}]{wanderley2023}
{Wanderley}, F., {Cunha}, K., {Souto}, D., {et~al.} 2023, \href{http://dx.doi.org/10.3847/1538-4357/acd4bd}{\JournalTitle{\apj}, 951, 90}

\bibitem[{{Wang} \& {Chen}(2019)}]{wang2019}
{Wang}, S., \& {Chen}, X. 2019, \href{http://dx.doi.org/10.3847/1538-4357/ab1c61}{\JournalTitle{\apj}, 877, 116}

\bibitem[{{Wilson} {et~al.}(2019){Wilson}, {Hearty}, {Skrutskie}, {Majewski}, {Holtzman}, {Eisenstein}, {Gunn}, {Blank}, {Henderson}, {Smee}, {Nelson}, {Nidever}, {Arns}, {Barkhouser}, {Barr}, {Beland}, {Bershady}, {Blanton}, {Brunner}, {Burton}, {Carey}, {Carr}, {Colque}, {Crane}, {Damke}, {Davidson}, {Dean}, {Di Mille}, {Don}, {Ebelke}, {Evans}, {Fitzgerald}, {Gillespie}, {Hall}, {Harding}, {Harding}, {Hammond}, {Hancock}, {Harrison}, {Hope}, {Horne}, {Karakla}, {Lam}, {Leger}, {MacDonald}, {Maseman}, {Matsunari}, {Melton}, {Mitcheltree}, {O'Brien}, {O'Connell}, {Patten}, {Richardson}, {Rieke}, {Rieke}, {Roman-Lopes}, {Schiavon}, {Sobeck}, {Stolberg}, {Stoll}, {Tembe}, {Trujillo}, {Uomoto}, {Vernieri}, {Walker}, {Weinberg}, {Young}, {Anthony-Brumfield}, {Bizyaev}, {Breslauer}, {De Lee}, {Downey}, {Halverson}, {Huehnerhoff}, {Klaene}, {Leon}, {Long}, {Mahadevan}, {Malanushenko}, {Nguyen}, {Owen}, {S{\'a}nchez-Gallego}, {Sayres}, {Shane}, {Shectman}, {Shetrone}, {Skinner}, {Stauffer}, \& {Zhao}}]{wilson2019}
{Wilson}, J.~C., {Hearty}, F.~R., {Skrutskie}, M.~F., {et~al.} 2019, \href{http://dx.doi.org/10.1088/1538-3873/ab0075}{\JournalTitle{\pasp}, 131, 055001}

\bibitem[{{Wright} {et~al.}(2011){Wright}, {Drake}, {Mamajek}, \& {Henry}}]{wright2011}
{Wright}, N.~J., {Drake}, J.~J., {Mamajek}, E.~E., \& {Henry}, G.~W. 2011, \href{http://dx.doi.org/10.1088/0004-637X/743/1/48}{\JournalTitle{\apj}, 743, 48}

\bibitem[{{Yang} \& {Johns-Krull}(2011)}]{yang2011}
{Yang}, H., \& {Johns-Krull}, C.~M. 2011, \href{http://dx.doi.org/10.1088/0004-637X/729/2/83}{\JournalTitle{\apj}, 729, 83}

\bibitem[{{Yang} {et~al.}(2008){Yang}, {Johns-Krull}, \& {Valenti}}]{yang2008}
{Yang}, H., {Johns-Krull}, C.~M., \& {Valenti}, J.~A. 2008, \href{http://dx.doi.org/10.1088/0004-6256/136/6/2286}{\JournalTitle{\aj}, 136, 2286}

\bibitem[{{Zacharias} {et~al.}(2012){Zacharias}, {Finch}, {Girard}, {Henden}, {Bartlett}, {Monet}, \& {Zacharias}}]{zacharias2012_vmag}
{Zacharias}, N., {Finch}, C.~T., {Girard}, T.~M., {et~al.} 2012, \JournalTitle{VizieR Online Data Catalog}, I/322A

\end{thebibliography}

\end{document}